\documentclass[12pt,preprint]{aastex}
\usepackage{graphicx}

\shorttitle{Multiple SGBs in eight GCs} 
\shortauthors{Piotto et al.} 
\usepackage{ulem}

\begin{document}

\title{ \textit{Hubble Space Telescope} reveals multiple Sub-Giant Branch in
  eight Globular Clusters\footnote{           Based on observations with  the
                               NASA/ESA {\it Hubble Space Telescope},
                               obtained at  the Space Telescope Science
                               Institute,  which is operated by AURA, Inc.,
                               under NASA contract NAS 5-26555.}}

\author{
G.\ Piotto\altaffilmark{2,3},
A.\ P. \,Milone\altaffilmark{4,5},
J.\ Anderson\altaffilmark{6}, 
L.\ R. \,Bedin\altaffilmark{3},
A.\ Bellini\altaffilmark{6},
S.\ Cassisi\altaffilmark{7}, 
A.\ F. \,Marino\altaffilmark{8},
A.\ Aparicio\altaffilmark{4,5},
V.\ Nascimbeni\altaffilmark{2}
 }

\altaffiltext{2}{Dipartimento  di   Astronomia,  Universit\`a  di Padova,
           Vicolo dell'Osservatorio 3, Padova I-35122, Italy;
           giampaolo.piotto@unipd.it }

\altaffiltext{3}{INAF-Osservatorio Astronomico di Padova, Vicolo
             dell'Osservatorio 5, Padova I-35122, Italy;
             luigi.bedin@oapd.inaf.it}

\altaffiltext{4}{Instituto de Astrof\`\i sica de Canarias, E-38200 La
              Laguna, Tenerife,  Canary Islands, Spain [milone,
                aparicio]@iac.es} 

\altaffiltext{5}{Department of Astrophysics, University of La Laguna,
           E-38200 La Laguna, Tenerife, Canary Islands, Spain}

\altaffiltext{6}{Space Telescope Science Institute, 3800 San Martin
  Drive, Baltimore, MD 21218; [jayander,bellini]@stsci.edu}

\altaffiltext{7}{INAF-Osservatorio Astronomico di Collurania, via Mentore
           Maggini, I-64100 Teramo, Italy cassisi@oa-teramo.inaf.it} 

\altaffiltext{8}{Max Planck Institute for Astrophysics, Postfach 1317,
	   D-85741 Garching, Germany; amarino@MPA-Garching.MPG.DE}

\begin{abstract}
In the last few years  many globular clusters (GCs) have revealed complex 
color-magnitude diagrams, with the presence of multiple main sequences (MSs), 
broaden or multiple  sub-giant branches (SGBs) and MS turn offs, and broad or 
split red giant branches (RGBs).  After a  careful correction for differential 
reddening, high accuracy photometry with the  Hubble Space Telescope presented 
in this paper reveals  a  broadened or even split SGB in five additional Milky  Way  GCs: NGC 362, NGC 5286,  NGC 6656,  NGC 6715, and  NGC 7089.  In addition, we 
confirm (with new and archival HST data) the presence of a split SGB in 47Tuc, 
NGC 1851, and NGC 6388.  The fraction of faint SGB stars with respect to the entire 
SGB population varies from one cluster to another and ranges from $\sim$0.03 
for NGC 362 to $\sim$0.50 for NGC 6715.  The average magnitude difference 
between the bright SGB and the faint SGB is almost the same at different 
wavelengths.  This peculiarity is consistent with the presence of two groups 
of stars with either an  age difference  of about 1-2 Gyrs,  or a significant 
difference in their overall C+N+O content.
\end{abstract}


\section{Introduction}
\label{introduction}
Recent photometric studies have revealed unexpectedly complex color-magnitude 
diagrams (CMDs) in many Globular Clusters (GCs), indicating that these stellar 
systems are not as simple as we have been assuming for decades (see Piotto 2009 
for a recent review).

$\omega$ Centauri is the most famous example of a GC hosting complex multiple 
stellar populations.  
It has been known since  the late nineties that its CMD shows  multiple 
red-giant branches (RGBs, Lee et al.\ 1999, Pancino et al.\ 2000), 
sub-giant branches (SGBs e. g.\  Bellini et al.\ 2010) and  multiple main
sequences (MSs, Anderson et al.\ 1997, Bedin et al.\ 2004).  

To date, multiple or broad RGBs have been observed in nearly all the GCs 
that have been observed with good signal-to-noise in the appropriate 
photometric bands (e. g.\ Marino et al.\ 2008, Yong et al.\ 2008, 
Lee et al.\ 2009, Lind et al.\ 2011).  Recent studies also suggest that 
multimodal MSs could be quite common among GCs.  In addition to the 
spectacular case of NGC 2808, which shows three  distinct MSs 
(Piotto et al.\ 2007), double and triple  MSs have been observed in 
several GCs, including NGC 104, NGC 6752, and NGC 6397 
(Anderson et al.\ 2009, Milone et al.\ 2010, 2012a,b) and have
been associated to stellar generations with a different content of
helium and light-elements (e. g.\ Norris 2004, Piotto et al.\ 2005,
D'Antona et al.\ 2005).  

Also the CMDs of many stellar clusters in the Large and the Small
Magellanic Cloud (LMC,  SMC) are not  consistent with single stellar
populations (Bertelli et al.\  2003, Mackey  et al.\ 2008).  Milone  
et al.\  (2009a) analyzed sixteen intermediate age clusters in the LMC 
and found that at least eleven of them (i.e.\ about the 70 \% of the 
whole sample) exhibit a  wide spread or a split around the main sequence 
turn-off (MSTO), which is consistent with the presence of multiple or 
prolonged star formation episodes.

Spectroscopic studies have long shown that most of the GCs exhibit 
   some correlations and anticorrelations among light-element abundances (such as 
the Na-O anticorrelation, Kraft et al.\ 1979, 1994, Ramirez \& Cohen 2002).
The fact that light-element variations have also 
been observed among unevolved stars (e.g.\ Cannon et al.\ 1998, Gratton et al.\
 2001) indicates that they have a primordial origin (see Gratton et al.\ 2004 for a review).

However, it must be noted that
spectroscopic analysis is necessarily limited to a small sample of
(bright) stars, and therefore allows us to identify stellar generations
 which may constitute only a small fraction of the cluster populations. Often spectroscopy investigations cannot follow multiple stellar generations in all evolving sequences (in particular along the main sequence and the white dwarf cooling sequence). Information on multiple stellar generations is maximized when spectroscopic and photometric data can be used together. Otherwise, we  must rely on photometry for a complete analysis of multiple populations (relative fraction, radial distribution, main chemical properties), which manifest themselves in different ways, in different clusters and different evolutionary parts of the CMD.

The multiple populations in NGC 1851 and NGC 104 manifest themselves
most prominently in terms of a splitting of the SGB (Milone et al.\  2008, 2012a,
Anderson  et al.\ 2009).  NGC 6388 also shows hints of such a splitting 
as well (Piotto 2008 and Moretti et al.\ 2009).  These SGB splits have 
been interpreted as indicating two groups of stars with either an age 
difference  of about  1-2 Gyrs, or a significant difference in their 
overall C+N+O content (Cassisi et al.\ 2008, Ventura et al.\ 2009, 
Di Criscienzo et al.\ 2010). 

In this paper we present detection of the broad (and most likely split) 
SGBs in five additional  Milky Way GCs: NGC 362, NGC 5286,  NGC 6656,  
NGC 6715, and  NGC 7089, and confirm the presence of a double SGB in 
NGC 6388, NGC 104, and NGC 1851.
In a companion paper based on the spectroscopy of stars selected 
from the CMDs published in this work, we have found that the
 two SGBs of NGC 6656 (M22) are made of stars with a different content of
 iron, $s$-process elements, and C+N+O (Marino et al.\ 2012). 
A bimodality in $s$-elements has been also observed along the RGB of NGC 1851 and associated with the double SGB of this cluster (e.g.\ Yong et al.\ 2008, Villanova et al.\ 2010, Lardo et al.\ 2012).

This paper is organized as follows.  In Sect.~\ref{data} we describe 
the data set, the photometric reduction, the zero point calibrations, the selection of stars with high-quality photometry, and the differential-reddening correction.  The initial
detection of the  SGB-broadenings is presented  in Sect.~3.  We then proceed to
study the SGBs in increasing detail.  We first use new and archival 
material: $i)$ to derive proper motions to exclude field 
objects (Sect.~4) and $ii)$ to provide confirmation of the SGB-broadenings
from several independent data-sets (Sect.~5).  Section~6 then examines
these SGBs through a variety of bandpasses.  Finally, in Sect.~7, 
for each cluster we compare the fractions of stars in the different 
SGBs with the distribution of stars along the HB in hopes of identifying the
same populations in the different evolutionary
branches (Sect.~8).  Section~9 contains 
the summary of results and an examination of open questions.

\section{Observations and data reduction}
\label{data}
We have used  images taken with the Wide
Field Planetary Camera 2 (WFPC2), the Wide Field Channel of the
Advanced Camera for Surveys (WFC/ACS), and the ultraviolet and
infrared channels of the Wide Field Camera 3 (UVIS and IR/WFC3) on board the
Hubble Space Telescope (\textit{HST}). 
These datasets include both proprietary images (GO11233, GO11739,
GO12311, PI Piotto) and data  that we retrieved from the STScI archive.   In 
 Table~1 we  give a  summary of  the 
adopted data sets indicating:\ the cluster  ID, the epoch
of  the  observation,  the  number  and the  duration  of  the  single
exposures, the filters, the program ID, and the program PI.

The astrometry and photometry of WFPC2 images have been carried out
for each chip, filter and epoch independently, by adopting the method
described by Anderson \& King (2000) based on the
effective-point-spread function fitting. We corrected for the
$34^{\rm th}$ row error in \textit{HST}'s WFPC2 cameras (see Anderson \& King \
1999 for details) and used the distortion solution 
as given by Anderson \& King \ (2003).  Photometric calibration has been done
according to the Holtzman et al.\ (1995) Vega-mag flight system for the WFPC2
camera.

The astro-photometric reduction of ACS/WFC data was done with the
software tools described in details by Anderson et al.\ (2008).
It consists in a package that analyzes all the exposures of each
cluster simultaneously in order to generate a single list of stars for
each field. 
Stars are measured in each exposure with a library PSF that is
``perturbed'' to better match the PSF in that exposure.
This routine was  designed to work well in  both crowded and uncrowded
fields and it is able to detect almost every star that can be detected
by eye. It takes advantage of the many independent dithered pointings of
each image and  the knowledge of the PSF  to avoid including artifacts
in the  list. Calibration of  ACS photometry into the  Vega-mag system
has been performed following recipes in Bedin et al.\ (2005) and using
the zero points given in Sirianni et al.\ (2005).
For ACS/WFC data coming from the GC Treasury program GO10775 (PI:\
Sarajedini) we simply used the calibrated catalogs produced by
Anderson et al.\ (2008, see also Sarajedini et al.\ 2007) using the
same procedures.

We measured star positions and fluxes on the WFC3
  images with a software mostly based on img2xym\_WFI (Anderson et
  al.\ 2006). Details on this program will be given in a stand-alone
  paper. Star positions and fluxes have been corrected for geometric
  distortion and pixel-area using the solutions provided by Bellini,
  Anderson \& Bedin (2011) for WFC3/UVIS and Anderson et al.\ (in preparation) for WFC3/IR.
%
%
\begin{table*}[ht!]
\scriptsize{
\begin{tabular}{llccccr}
\hline\hline  ID & DATE & N$\times$EXPTIME & FILT & INSTRUMENT & PROGRAM & PI  \\
\hline
\hline
 NGC 362  & Dec 04 2003                      & 2$\times$110s$+$2$\times$120s   & F625W  & ACS/WFC &  10005 & Lewin      \\
          & Dec 04 2003                      & 4$\times$340s                   & F435W  & ACS/WFC &  10005 & Lewin      \\
          & Sep 30 2005                      & 3$\times$70s$+$20$\times$340s   & F435W  & ACS/WFC &  10615 & Lewin      \\
          & Jun 02 2006                      & 10s$+$4$\times$150s             & F606W  & ACS/WFC &  10775 & Sarajedini \\
          & Jun 02 2006                      & 10s$+$4$\times$170s             & F814W  & ACS/WFC &  10775 & Sarajedini \\
 NGC 1851 & Nov 11-Dec 27 2010, Jan 2 2011  & 14$\times$1280s            & F275W  & WFC3/UVIS& 12311 & Piotto     \\
          & Nov 11-Dec 27 2010, Jan 2 2011  &  7$\times$100s             & F814W  & WFC3/UVIS& 12311 & Piotto     \\
          & Apr 10 1996                      & 4$\times$900s                   & F336W  & WFPC2   &  5696  & Bohlin     \\
          & May 01 2006                      & 20s$+$5$\times$350s             & F606W  & ACS/WFC & 10775  & Sarajedini \\		
          & May 01 2006                      & 20s$+$5$\times$360s             & F814W  & ACS/WFC & 10775  & Sarajedini \\		
 NGC 5286 & Nov 07 1997                      & 10s$+$4$\times$140s             & F555W  & WFPC2   &  6779  & Gebhardt   \\
          & Nov 07 1997                      &  7s$+$3$\times$140s             & F814W  & WFPC2   &  6779  & Gebhardt   \\
          & Mar 03 2006                      & 30s$+$5$\times$350s             & F606W  & ACS/WFC & 10775  & Sarajedini \\		
          & Mar 03 2006                      & 30s$+$5$\times$360s             & F814W  & ACS/WFC & 10775  & Sarajedini \\		
          & Mar 24 2009                      & 60s$+$3$\times$700s             & F336W  & WFPC2   & 11975  & Ferraro    \\
          & Mar 24 2009                      &  2s$+$3$\times$100s             & F336W  & WFPC2   & 11975  & Ferraro    \\
 NGC 6388 & from Sep 04 2003 to Jun 26 2004  &  6$\times$11s                   & F435W  & ACS/WFC &  9821  & Pritzl     \\
          & from Sep 02 2003 to Jun 23 2004  &  6$\times$7s                    & F555W  & ACS/WFC &  9821  & Pritzl     \\
          & from Oct 02 2003 to Jun 23 2004  &  6$\times$3s                    & F814W  & ACS/WFC &  9821  & Pritzl     \\
          & Apr 06 2006                      & 40s$+$5$\times$340s             & F606W  & ACS/WFC & 10775  & Sarajedini \\		
          & Apr 06 2006                      & 40s$+$5$\times$350s             & F814W  & ACS/WFC & 10775  & Sarajedini \\		
          & Apr 22-23 and May 17 2008        & 11$\times$700s$+$11$\times$800s & F450W  & WFPC2   & 11233  & Piotto     \\
          & Apr 22-23 and May 17 2008        & 11$\times$500s                  & F814W  & WFPC2   & 11233  & Piotto     \\
          & Jun 30 and Jul 3 2010            &  6$\times$800s                  & F390W  & WFC3/UVIS& 11739 & Piotto     \\
          & Jun 30 2010                      &  6$\times$199s$+$4$\times$249s  & F160W  & WFC3/IR & 11739  & Piotto     \\
 NGC 6656 & Jun 06 2000                      & 12$\times$26s 	               & F555W  & WFPC2   &  8174  & van Altena \\
          & Jun 06 2000                      & 11$\times$26s                   & F785LP & WFPC2   &  8174  & van Altena \\
          & Apr 01 2006                      &  3s$+$4$\times$55s              & F606W  & ACS/WFC & 10775  & Sarajedini \\		
          & Apr 01 2006                      &  3s$+$4$\times$65s              & F814W  & ACS/WFC & 10775  & Sarajedini \\		
          & Apr 29 2008                      &  9$\times$350s                  & F450W  & WFPC2   & 11233  & Piotto     \\
          & Apr 29 2008                      &  9$\times$100s                  & F814W  & WFPC2   & 11233  & Piotto     \\
          & Apr 22 2009                      &  3$\times$10s                   & F555W  & WFPC2   & 11975  & Ferraro    \\
          & Apr 22 2009                      &  3$\times$350s                  & F336W  & WFPC2   & 11975  & Ferraro    \\
          & Sep 23 2010 and Mar 18 2011      &  9$\times$812s                  & F275W  & WFC3/UVIS&12311  & Piotto     \\
          & Sep 23 2010 and Mar 18 2011      &  4$\times$50                    & F814W  & WFC3/UVIS&12311  & Piotto     \\
 NGC 6715 & Maj 25 2006                      & 2$\times$30s$+$10$\times$340s   & F606W  & ACS/WFC & 10775  & Sarajedini \\		
          & Maj 25 2006                      & 2$\times$30s$+$10$\times$350s   & F814W  & ACS/WFC & 10775  & Sarajedini \\		
          & Jun 20-22-24 2007                &  9$\times$700s$+$9$\times$800s  & F450W  & WFPC2   & 10922  & Piotto     \\
          & Jun 20-22-24 2007                &  9$\times$500s                  & F814W  & WFPC2   & 10922  & Piotto     \\
          & Aug 30 1999                      &  5$\times$300s$+$350            & F555W  & WFPC2   &  6701  & Ibata     \\
          & Aug 30 1999                      &  260s$+$5$\times$300s           & F814W  & WFPC2   &  6701  & Ibata     \\
 NGC 7089 & Apr 16 2006                      & 20s$+$5$\times$340s             & F606W  & ACS/WFC & 10775  & Sarajedini \\
          & Apr 16 2006                      & 20s$+$4$\times$340s             & F814W  & ACS/WFC & 10775  & Sarajedini \\
          & Aug 22 2000                      & 600s$+$2$\times$700s            & F250W  & WFPC2   &  8709  & Ferraro    \\
          & Aug 22 2000                      & 2$\times$50s$+$500s$+$2$\times$600s & F336W  & WFPC2   &  8709  & Ferraro    \\
          & Jun 27 2008                      & 10$\times$700s                  & F450W  & WFPC2   & 11233  & Piotto     \\
          & Jun 27 2008                      & 230s$+$9$\times$200s            & F450W  & WFPC2   & 11233  & Piotto     \\
\hline
\end{tabular}
\label{tabelladati}
\caption{Description of the \textit{HST} archive data sets used in this paper. 
}}
\end{table*}
\subsection{Selection of the star sample}
\label{selezione}
Individual stars on  the   
images  may  be  poorly   measured  for  several
reasons: crowding by nearby  neighbors, contamination from cosmic rays,
or image artifacts, such as bad  pixels, or diffraction spikes.  The goal
of the present work is to clearly identify multiple populations, which
often manifest themselves as a splitting of 0.02 magnitudes or less
in the cluster sequences.
For this purpose  we need to select  the best-measured stars  in the field
(i.e.\ those  with the  lowest random and  systematic errors),  but we
also need  to have  a large  enough statistical sample  to be  able to
identify secondary sequences that may have 
a much smaller fraction of stars 
than the primary ones.

The software described in Anderson  et  al.\  (2008) provides  very
valuable tools to reach this  goal.  In addition to the stellar fluxes
and positions it  generates a number of parameters that  can be used as
diagnostics   of   the   reliability  of   photometric   measurements.
These include a parameter, $q$, related to the quality of the fit of the
PSF model to the star's pixels,
the difference in position for each star as measured in different filters,
$rms_{\rm X, Y}$, the $rms$
of magnitudes measured in different  exposures, $rms_{\rm mag}$, and a
parameter, $o$, which quantifies how  much a star is  contaminated  from
neighboring stars flux (see Anderson et al.\ 2008 for the details).

To select the sample of stars with the best photometry, we adopted the
following  procedure, which is described  in  more  details  in  Sect.~2.1 
of  Milone et  al.\  (2009a).  Since the  $q$ and  the $rms$
parameters exhibit a trend with  the magnitude, we divided the stars
into bins  of 0.04  magnitude and calculated for each bin the median
of the $rms$ and $q$ parameters and the  $68.27^{\rm th}$ percentile 
(hereafter $\sigma$), by means of the iterative procedure  explained in 
Milone et al.\ (2009a).  We then added  to the  median  of each  bin four
times $\sigma$  and interpolated these points with a  spline. All the stars 
that lie below the spline were flagged as well-measured.  The $o$  parameter 
does not show  a clear trend with  magnitude, so we considered all the 
stars with $o<0.5$ as well-measured.
 In the cases of the most crowded
clusters (namely NGC 5286, NGC 6388, NGC 6715, and NGC 7089) we have limited our 
investigation
to stars with radial distance from the cluster center larger than 40 arcsec.

 The fraction $f$ of stars that pass all the other criteria of selection, with respect to the total 
number of stars changes from one cluster to the other, and strongly depends 
on the dataset, and less critically on 
the luminosity of the star. As an example, in the case of NGC 6388, which is one the most crowded GCs of this paper,
we have $f$=0.66 for the F606W ACS/WFC GO10775 images and
for stars brighter than $m_{\rm F606W}=23.0$. Specifically, 
$\sim$78\% and $\sim$77\% of the stars pass the criteria of selection based on the positions and magnitude $rms$ respectively,
$\sim$76\%  of stars pass the criteria based on the $q$ parameter, 
$\sim$78\% that based on $o$, and 66\% all the criteria above. 
In the case of NGC6656, the most sparse GC in our sample, 
for F606W ACS/WFC images from GO10775 we have
$f$=0.89 for stars with $m_{\rm F606W}<23.0$.

\subsection{Correction for differential reddening, and zero point spatial variations}
The foreground reddening of the GCs studied in this paper goes from
0.05 (NGC 362) to 0.37 (NGC 6388, Harris 1996, 2010).
Therefore, variations of reddening are expected within
the field of view of our target clusters.
Indeed, the CMDs in all clusters reveal that all the CMD sequences are
significantly broader than what would be expected from our photometric
precision, even taking into account the small unmodelable 
spatial changes of the PSF, which cause photometric 
zero-point variations usually smaller than 0.01 magnitude
(see Anderson et al.\ 2008 for details).

Correcting for both differential reddening and the residual spatial variations
of the photometric zero point represents a necessary step towards an
accurate analysis of the cluster sequences.  
To obtain these corrections, we applied
the method described in Milone et al.\ (2012c, Sect.~3).
Briefly, we  defined  the fiducial MS for the cluster and estimated,
for each star, how the observed stars in its vicinity may 
systematically lie  to the red or  the blue of  the fiducial sequence;
this systematic  
color 
and magnitude offset, measured along the
reddening line,  is indicative of the local differential reddening.

\begin{figure*}[ht!]
   \centering
  \epsscale{.7}
   \plotone{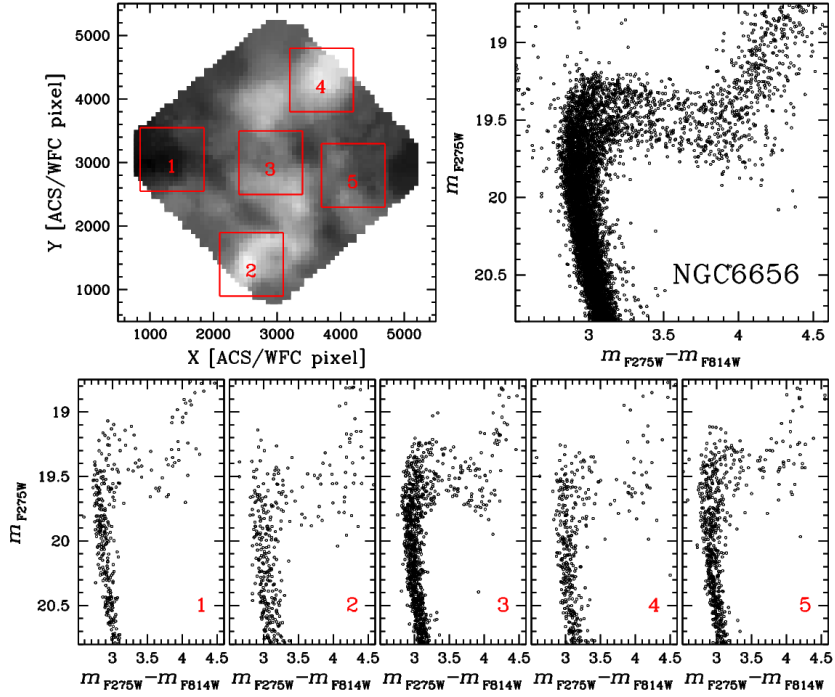}
      \caption{\textit{Upper-left:} map of differential reddening in
        the NGC 6656 field of view. The grey levels indicate the
        amplitude of differential reddening, with black and white
        corresponding to $E(B-V)$=$-$0.07 and $E(B-V)$=0.09 respectively.   
        \textit{Upper-right:} $m_{\rm F275W}$
        versus $m_{\rm F275W}-m_{\rm F814W}$ CMD of NGC 6656 corrected
        for differential reddening and zoomed around the SGB. 
        Lower panels show the
        original CMDs, without any differential-reddening correction,
        for stars in the five 1000$\times$1000 pixel regions
        indicated in the upper-left panel.} 
         \label{map6656}
   \end{figure*}

As an example,
in the upper-left panel of Fig.~\ref{map6656} we plot the map of differential reddening in the field of view of NGC 6656. The CMD corrected
for differential reddening is shown in the upper-right panel, while
the original CMDs of stars in five, distinct  1000$\times$1000 pixel  
regions are plotted in the lower panels. 
The fact that the double SGB is visible everywhere in
the field of view further demonstrates that it cannot be an artifact
introduced by differential reddening.

\begin{figure*}[ht!]
   \centering
  \epsscale{.99}
   \plotone{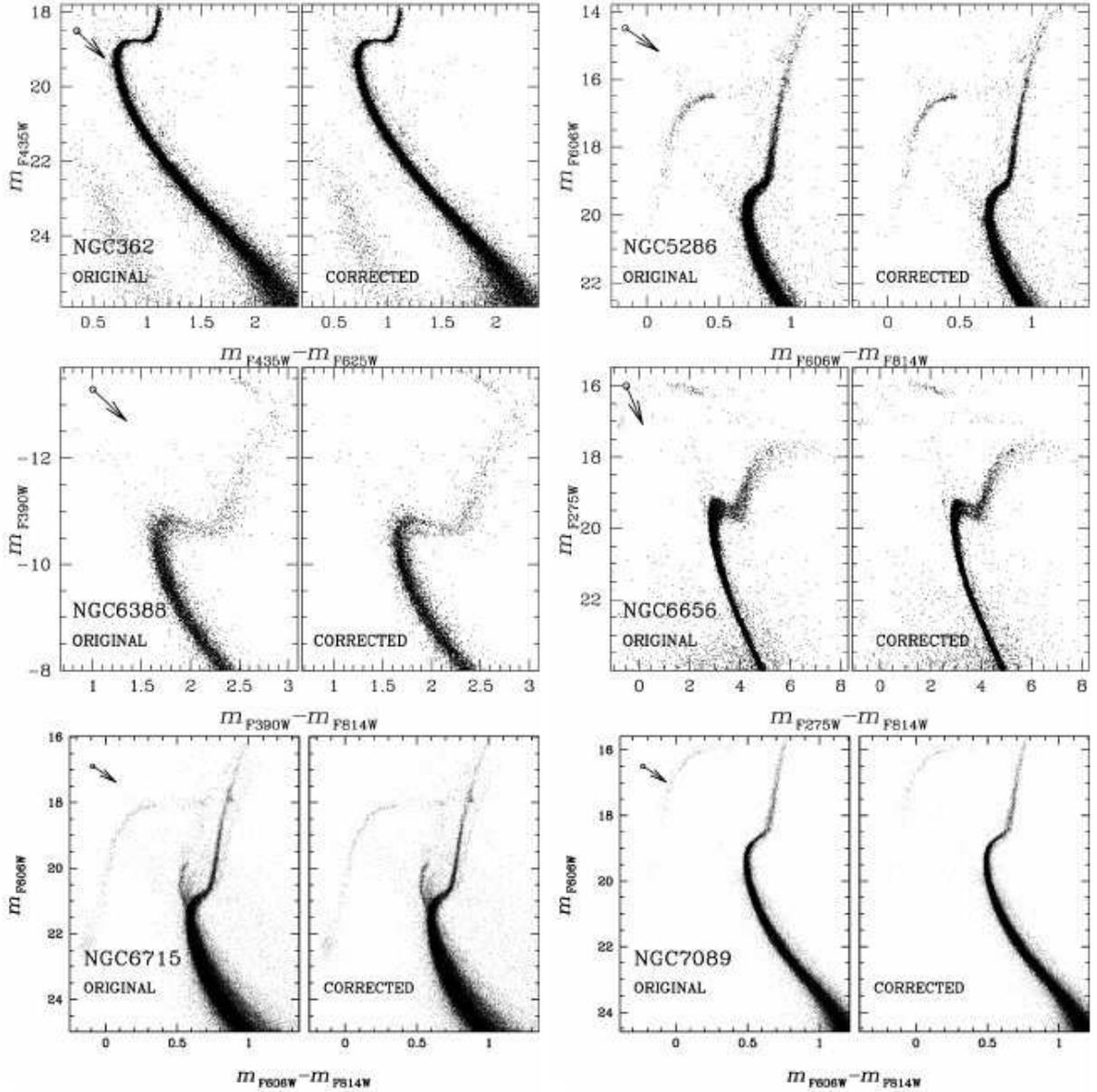}
      \caption{Comparison between the original CMD (left) and the CMD corrected for differential reddening (right), for six of our cluster sample. The arrows indicate the reddening direction in each CMD. }
         \label{corrNOTcorr}
   \end{figure*}

In Fig.~\ref{corrNOTcorr} we show some CMDs before and after our
differential-reddening correction for  six out of the eight clusters considered in this work.

%
\clearpage

\section{Multiple stellar populations along the SGB}
\label{SGB6}
\begin{figure}[hp!]
   \centering
  \epsscale{.89}
   \plotone{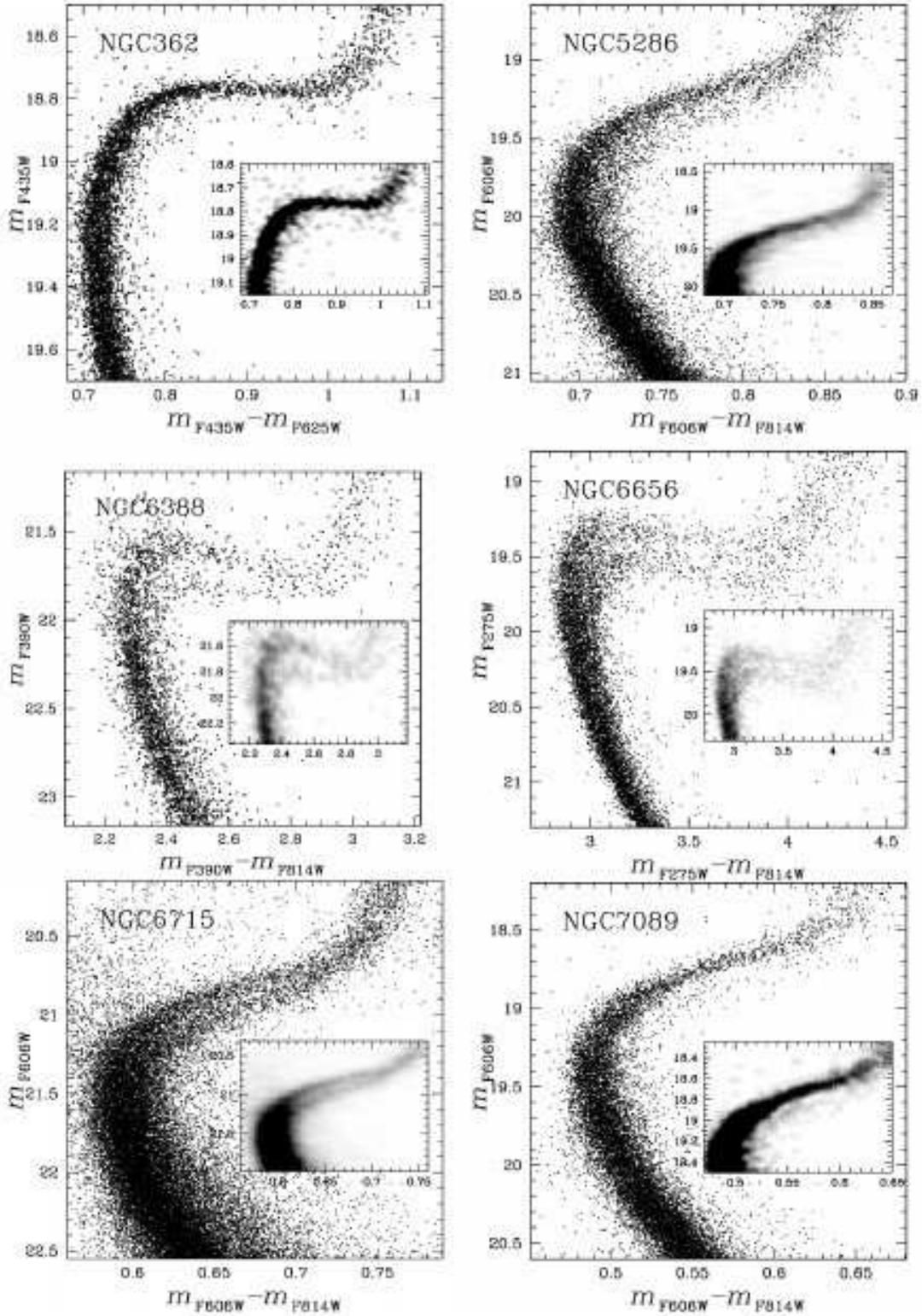}
      \caption{ 
	CMDs of six out of the eight GCs studied  in  this paper. Only  stars 
which successfully passed   all  the selection   criteria have  been
	plotted. In the inset,  we  show the  Hess diagram for  the 
	region around the SGB.
	  }
         \label{FIGSGB}
   \end{figure}

Fig.~\ref{FIGSGB} shows a zoom of the CMD region around the SGB for six
of the clusters studied in this paper.
Among the several data sets available for each cluster, we used here
the ones with the best photometric quality for SGB stars.
The most intriguing feature is already evident: the SGB is 
much broader than expected (indeed, in most cases split)
from photometric errors, which are smaller than 0.01 mag for these
selected stars.
Note that the fraction of stars in the fainter
 SGBs differs from cluster to cluster, as 
discussed in detail in Sect.~\ref{pratio}. 

\subsection{The multiple SGB of NGC 1851 and 47 Tucan\ae}
\label{SGB2}

Excluding the peculiar case of $\omega$\,Centauri, 
we now know from literature at least two additional GCs that show a SGB that is not consistent with a single stellar population\footnote{
    We note that the SGB of NGC 6441 is also suspected 
    to show two populations, but we postpone details and discussion 
    to a forthcoming paper dedicated to the two anomalous GCs NGC 6388 
    and NGC 6441 (Bellini et al.\ 2012, in prep.) which will be based on \textit{HST} program GO-11739.}.

The first is NGC 1851, which exhibits two clearly distinct SGB branches 
with a faint SGB containing about the 35 \% of the SGB stars (Milone et al.\ 2008, 2009b).  

The second is 47\,Tuc, for which  
evidence of multiple stellar populations along the SGB of 47\,Tuc came from the analysis of a large number of ACS/\textit{HST} images
by Anderson et al.\ (2009).
These authors found that the SGB exhibits a clear spread in luminosity, with at
least two distinct components: a brighter one with a 
significant spread in magnitude, and a second one about 0.05 mag
fainter in the  F475W band, containing about 10 \% of the stars.  

Milone et al.\ (2012a) found that the brighter SGB further splits into
two branches, and that the presence of two stellar populations is clearly visible in the MS, RGB, and the HB. 

\begin{figure}[hp!]
   \centering
  \epsscale{.75}
   \plotone{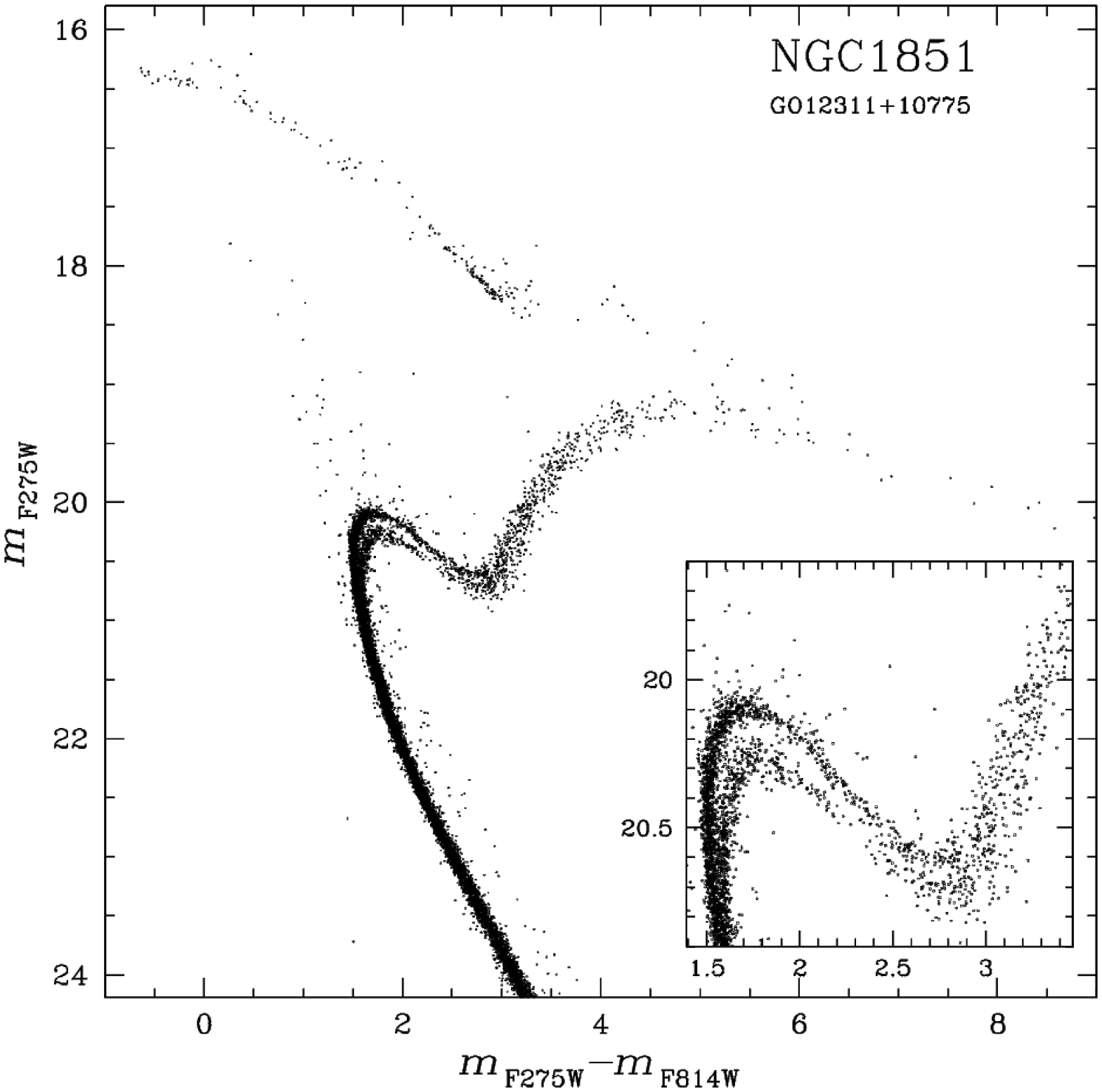}
      \caption{ 
	CMDs of NGC 1851 from ACS/WFC/F814W and WFC3/UVIS/F275W photometry. 
	The inset shows a zoom around the SGB.
	  }   
         \label{SGB1851}
   \end{figure}

These results have stimulated us to start a deeper investigation using the 
WFC3/UVIS camera:\ GO-12311 (PI Piotto).
The ${\it m}_{\rm F275W}$ versus ${\it m}_{\rm F275W}-{\it m}_{\rm
 F814W}$ CMD of NGC 1851, obtained combining these data and archive
ones, is shown in Fig.~\ref{SGB1851}. 
This CMD not only confirms the presence of a split SGB, but it
is so far the clearest bi-modal SGB in literature.

\begin{figure}[hp!]
   \centering
  \epsscale{.75}
   \plotone{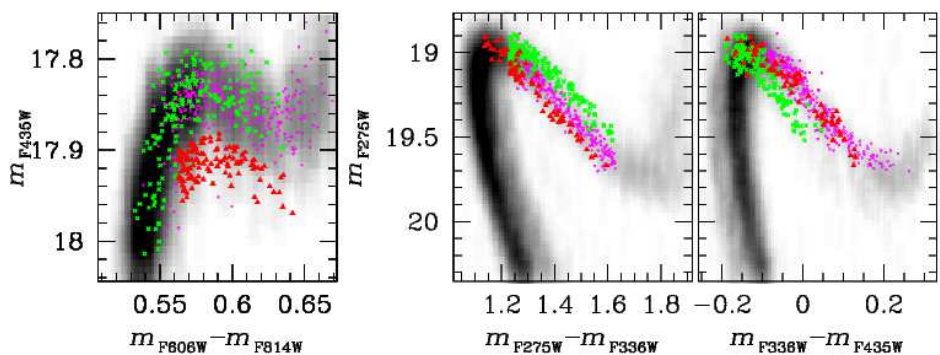}
      \caption{ 
        Reproduction of Fig.~12 from Milone et al.\ (2012a). We show
        the CMDs of 47 Tuc zoomed around the SGB. In the left panel we
        have colored fSGB stars red. The other two SGB components are
        clearly visible in the middle and the right CMDs.
	  }
         \label{SGBs47t}
   \end{figure}

In Fig.~\ref{SGBs47t} we reproduce 
Fig.~12 from Milone et al.\ (2012a), which shows the complex SGB of 47 Tuc.
In the ${\it m}_{\rm F435W}$ versus ${\it m}_{\rm F606W}-{\it m}_{\rm
  F814W}$ we see the faint SGB discovered by Anderson et al.\ (2009)
(these stars are colored red).
The ${\it m}_{\rm F275W}$ versus
${\it m}_{\rm F275W}-{\it m}_{\rm  F336W}$ and the ${\it m}_{\rm
  F275W}$ versus ${\it m}_{\rm F336W}-{\it m}_{\rm  F435W}$ CMDs are
plotted in the middle and the right panel and show the other two SGB
branches. 

Table~\ref{parameters} shows the 
relevant (for the present paper) parameters for the eight GCs with 
a significant SGB  broadening, and that we will discuss in the following. 
Most of them come from Harris (1996, updated as in 2010).
HBR  is the Horizontal Branch ratio HBR=(B-R)/(B+V+R), where B, R, and V are the numbers of blue HB, red HB and RR Lyrae (from Harris 1996, updated as in 2010). 
The HBR for NGC 6388 has been calculated from the CMD presented in here.
\begin{table}
\centering
\caption{
Main parameters for the GCs with split or spread SGB. }
\begin{tabular}{lccccccc}
\hline\hline  ID   &  $R_{\rm GC}$ & $E(B-V)$ & $(m-M)_{\rm V}$ & HBR &
$M_{\rm V}$ & [\rm Fe/H] \\
\hline
\hline
NGC  104  &  7.4 & 0.04 & 13.37 & $-$0.99 &  $-$9.42 & $-$0.76 \\
NGC  362  &  9.4 & 0.05 & 14.81 & $-$0.87 &  $-$8.41 & $-$1.16 \\
NGC 1851  & 16.7 & 0.02 & 15.47 & $-$0.36 &  $-$8.33 & $-$1.22 \\
NGC 5286  &  8.4 & 0.24 & 15.95 &    0.24 &  $-$8.61 & $-$1.67 \\
NGC 6388  &  3.2 & 0.37 & 16.14 & $-$0.63 &  $-$9.42 & $-$0.60 \\
NGC 6656  &  4.9 & 0.34 & 13.60 &    0.91 &  $-$8.50 & $-$1.76 \\
NGC 6715  & 19.2 & 0.15 & 17.61 &    0.75 &  $-$10.1 & $-$1.58 \\
NGC 7089  & 10.4 & 0.06 & 15.49 &    0.96 &  $-$9.02 & $-$1.62 \\
\hline
\label{parameters}
\end{tabular}
\end{table}

\section{Proper motions}
\label{PM}
The stellar fields analyzed in this paper are all located within few
arcminutes from the cluster centers.
As a consequence, in all the cases, the number of
field stars is negligible with respect to the cluster members.
However, while in NGC 5286, NGC 6388, NGC 6656, NGC 6715 both 
SGBs are well populated, and therefore well distinguishable from field stars, in
NGC 362 and NGC 7089 the faint SGB (fSGB) 
contains only a small fraction of the total number of SGB stars.
For these two clusters, the effect of the field contamination
is more relevant for the assessment of the significance of the fSGB.
One of the most obvious advantages of having more than one epoch
observations is the possibility to measure the proper motions of stars. 
In the following, we use proper motions to show that in all
clusters, both the fSGB and bright SGB (bSGB) belong to the cluster, and
cannot be associated to field-stars.

For NGC 362, NGC 5286, NGC 6388, NGC 6656, and  NGC 7089, we have used the 
proper-motion measurements from Milone et al.\ (2012c). 
For NGC 6715 we determined proper motions from GO6701 (hereafter epoch 1), 
GO10775 (epoch 2) and GO10922 (epoch 3) data.  To do this, we followed 
the procedure that has been used for the other clusters by 
McLaughlin et al.\ (2006), Bedin et al.\ (2009), and Anderson \& van der Marel (2010).

Briefly, we have adopted as reference frames the frames defined by 
Anderson et al.\ (2008) for GO10775 data set, which have $X$ increasing 
from east to west, and $Y$ from south to north.  
We corrected for distortion the coordinates of each star measured in each exposure,  and transformed the average position for each star in each of the three epochs into the master frame ($x^{\rm *,i}$, $y^{\rm *,i}$ i=1,2,3). 

To minimize the  effects of any  residual distortion, we applied local
transformations using a   local subsample of  stars. To  determine the
transformation for each star,  we selected the closest stars to the
target star as local reference stars.  The corresponding 55 pairs of 
coordinates in the two frames are used to determine  the  least-squares
linear transformation from one coordinate system to the other. Naturally,
we excluded the target star in its own transformation.

We used linear transformations in the form: 
\begin{equation}
\mathcal{X}^{\rm N,i}=a(x^{\rm N,i}-x_{0}) + b(y^{\rm N,i}-y_{0}) + \mathcal{X}_{0},
\end{equation}
\begin{equation}
\mathcal{Y}^{\rm N,i}=c(x^{\rm N,i}-x_{0}) + d(y^{\rm N,i}-y_{0}) + \mathcal{X}_{0}
\end{equation}
where  
$\mathcal{X}^{\rm N,i}, \mathcal{Y}^{\rm N,i}$, {\rm i=1,2,3}
are the coordinates of the target stars in the distortion-corrected master-frame, while 
$x^{\rm N,i}, y^{\rm N,i}$, {\rm i=1,2,3} 
are those in the non-reference frames,  
and the constants $a$,$b$,$c$,$d$, ($x_{0}$, $y_{0}$), and ($\mathcal{X}_{0}$,
$\mathcal{Y}_{0}$)  are  to  be  determined.

Finally we plotted the positions of the star along $\mathcal{X}$ and $\mathcal{Y}$ as a
 function of the epoch expressed in years and obtained
the  motion  components on the   plane  of the sky  by
standard, error-weighted least squares fit of a straight line to the points,
 allowing both the slope and the  zero point to vary. 
The best estimate of stellar proper motions (in pixels per year)
is given by the slope of the line multiplied
by the pixel-scale of the reference frame which is 49.7248 mas/pixel
(van der Marel et al.\ 2007). 

The selection of the reference stars is crucial for accurate transformations.
Therefore, we imposed that reference stars  must be unsaturated  in the 
images  of both epochs, and must pass the criteria of selection described 
in Sect.~\ref{data}.

We further note that the internal proper-motions of cluster stars are usually
negligible with respect to our errors\footnote{
     An exception is represented by the nearby GCs M~22, where the  
     internal cluster velocity dispersion is significantly larger 
     than the proper-motion errors but at least five times smaller 
     than field star proper motion.}.
Therefore, we can refer our motions to the bulk of cluster members 
motion.  In order to select a pure
sample of cluster members, we started by identifying all the stars  
that, on the basis of their position in the CMD, are probable cluster 
members and obtained a crude proper-motion estimate by using a local 
transformation based on this sample.   Then we have iteratively
excluded  from this  list all the stars that did not have cluster-like 
proper motions (proper motion $>2.5 \sigma$, where $\sigma$ is the 
proper-motion dispersion of cluster members), despite their proximity
to the cluster sequences.  This ensures that the relative motions of 
cluster stars  should  be, on average, zero.

\begin{figure}[ht!]
   \centering
  \epsscale{.79} 
   \plotone{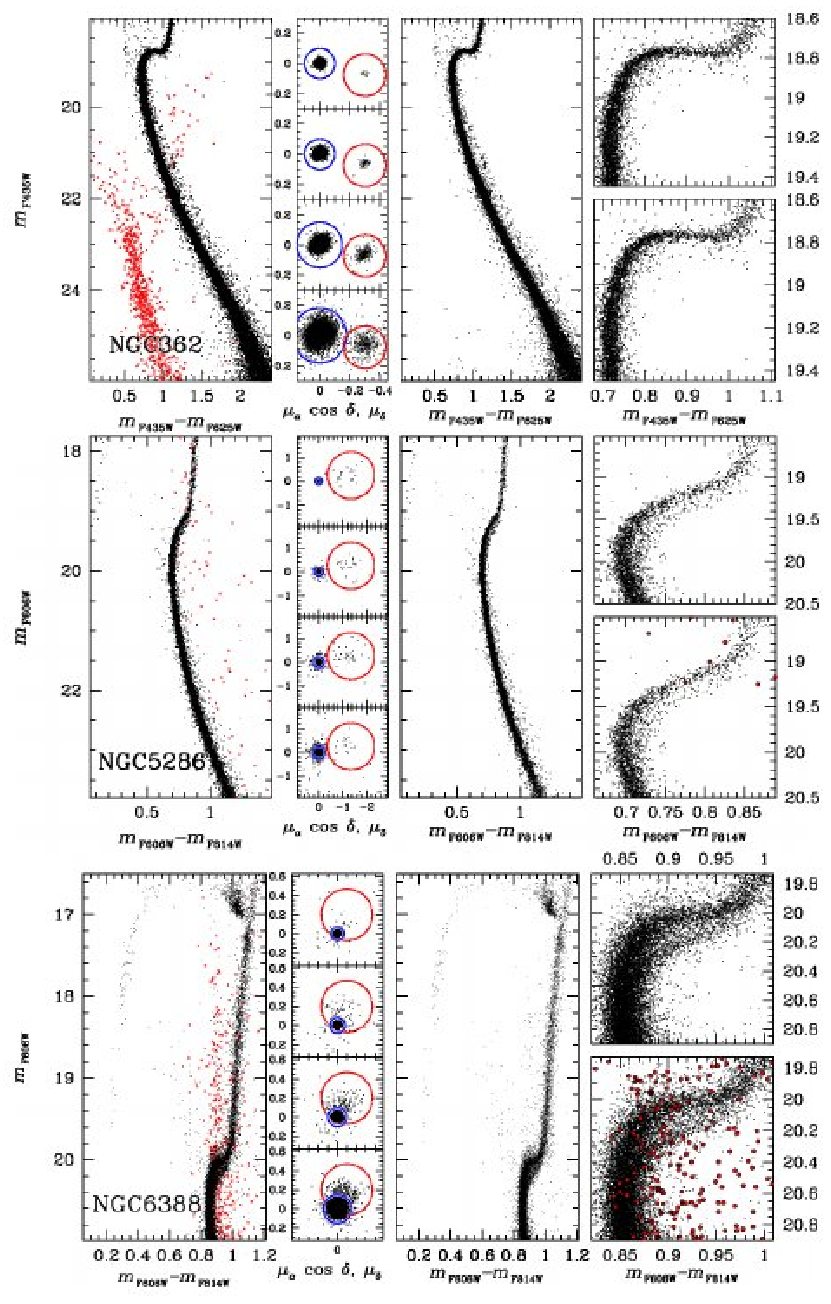}
      \caption{ 
\textit{Leftmost column}: F606W vs.\ F606W$-$F814W CMD (F435W vs.\ F435W-F625W in the case of NGC 362) for all stars
        with available proper motions.
	\textit{Second column}: Proper-motion diagrams of the stars 
                                 in the left panels, in intervals of 1.4 mag. 
        \textit{Third column}: The proper-motion selected CMD of cluster members
(all stars within the blue circles).
	\textit{Rightmost column}: Zoom of the proper-motion selected CMD 
                                    around the SGB.  In this figure we show 
                                    the cases of NGC 362, NGC 5286, NGC 6388.
	  }
         \label{PM_A}
   \end{figure}

\begin{figure}[ht!]
   \centering
  \epsscale{.79} 
   \plotone{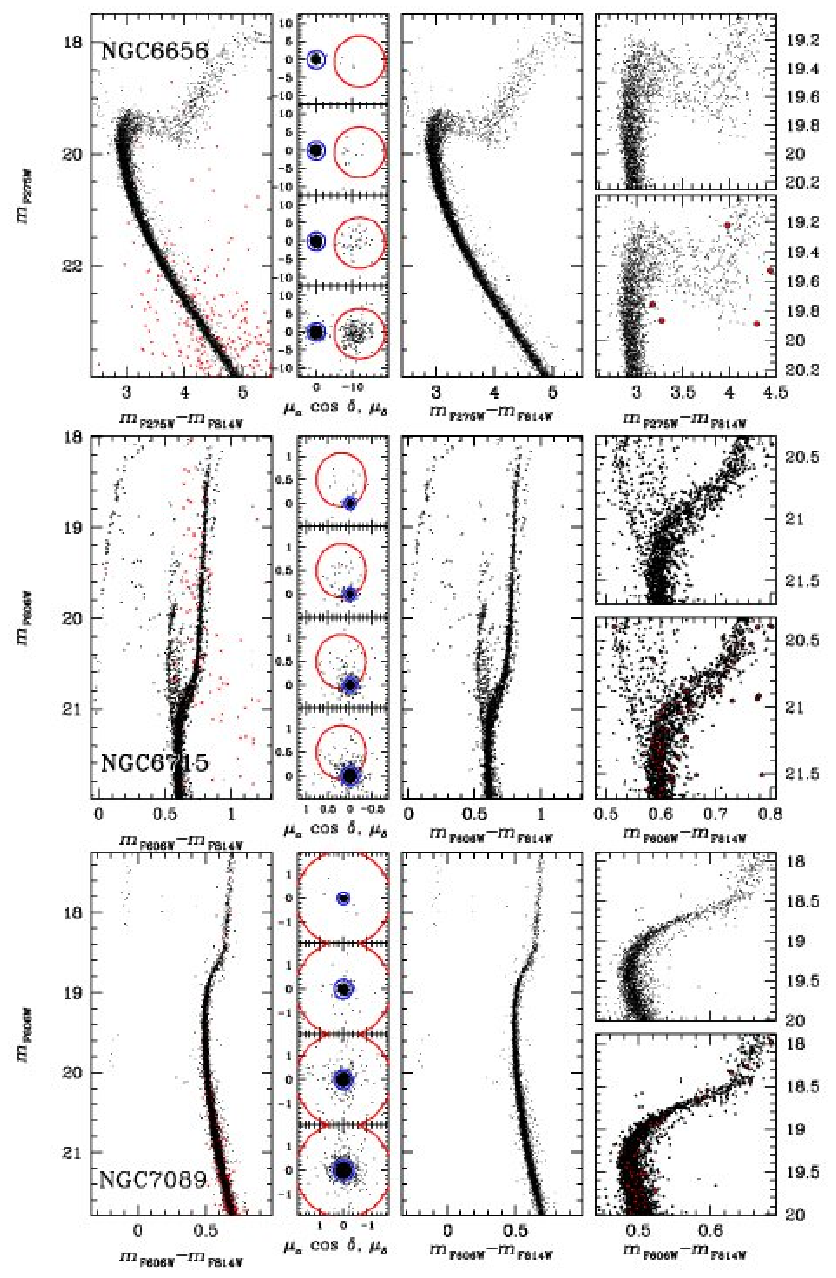}
      \caption{ As in 
Fig.~\ref{PM_A} for NGC 6656, NGC 6715 and NGC 7089.
	  }
         \label{PM_B}
   \end{figure}

Results (and proper-motion-selection criteria) are shown in 
Figs~\ref{PM_A}-\ref{PM_B}.  In the left panel we show the CMD for 
all the stars for which proper-motion measurements are available.
The second column of the panels shows the proper-motion diagrams 
of the stars for four different magnitude intervals.  The blue and 
red circles are drawn in order to select two groups of stars that have 
member-like and field stars-like motions, respectively. The red circle 
size has been chosen by eye with the criterion of rejecting the most 
evident field stars that are plotted in red in the left panel CMD. 

It should be noted that in our study it is more important to have a 
pure cluster sample than a complete one.  Therefore, we fixed the 
radius of  blue circles at 2.5 $\sigma$, where $\sigma$ is the proper-motion 
dispersion in each dimension. If we assume that proper motions of 
cluster stars follow a bivariate Gaussian distribution, each circle 
should include 95 \% of the cluster members in each magnitude interval. 
The third column panels show the CMD of our chosen cluster stars, and the 
rightmost upper panel is a blowup of the CMD region around the SGB. 
In the rightmost lower panel we marked in black the location of stars
with cluster like motion, and in red field stars.  
In all cases, the stars  in both the fSGB and bSGB regions have cluster-like proper motion and cannot be attributed to a field population. 

\section{Confirming the split/spread SGB with independent datasets}
\label{confirm}
\begin{figure*}[ht!]
   \centering
  \epsscale{.99} 
   \plotone{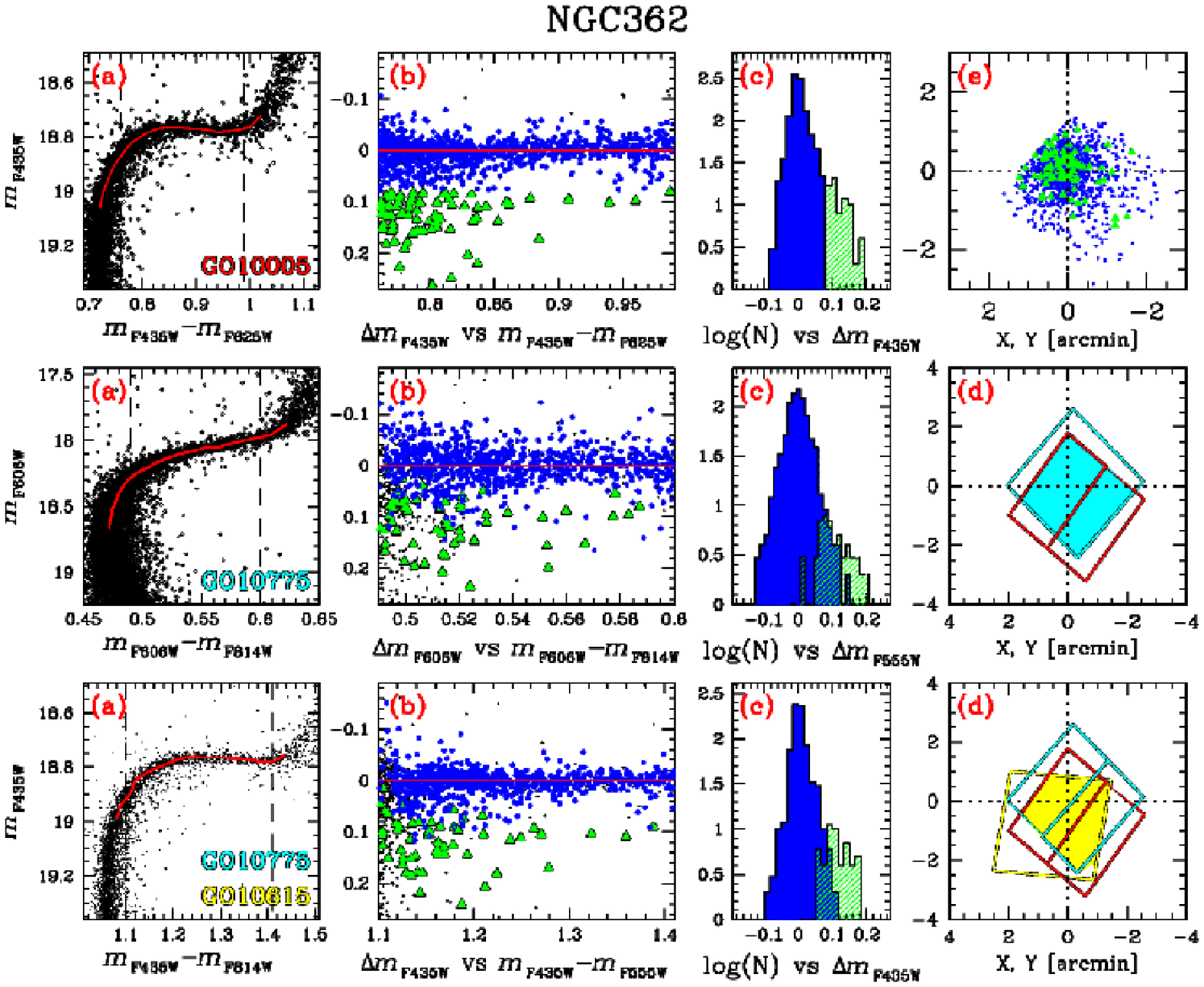}
      \caption{ \textit{Panel a:} Zoom of the CMD of NGC 362 around
        the SGB. \textit{Panel b:} Rectified SGB 
        with the groups of bSGB and fSGB stars colored in blue and
        green respectively. \textit{Panel c:} Distribution of the magnitude
        residuals ($\Delta$ m). \textit{Panel d:} Footprint of the
        data sets. The stars used in this analysis belong to the shadowed
        area.  \textit{Panel e} shows the spatial distribution for the
        bSGB and fSGB stars.
        We used the same color code for the dataset name in the panel a and its footprint in panel d. 
	  }
         \label{confirmSGB362}
   \end{figure*}

\begin{figure}[ht!]
   \centering
  \epsscale{.99} 
   \plotone{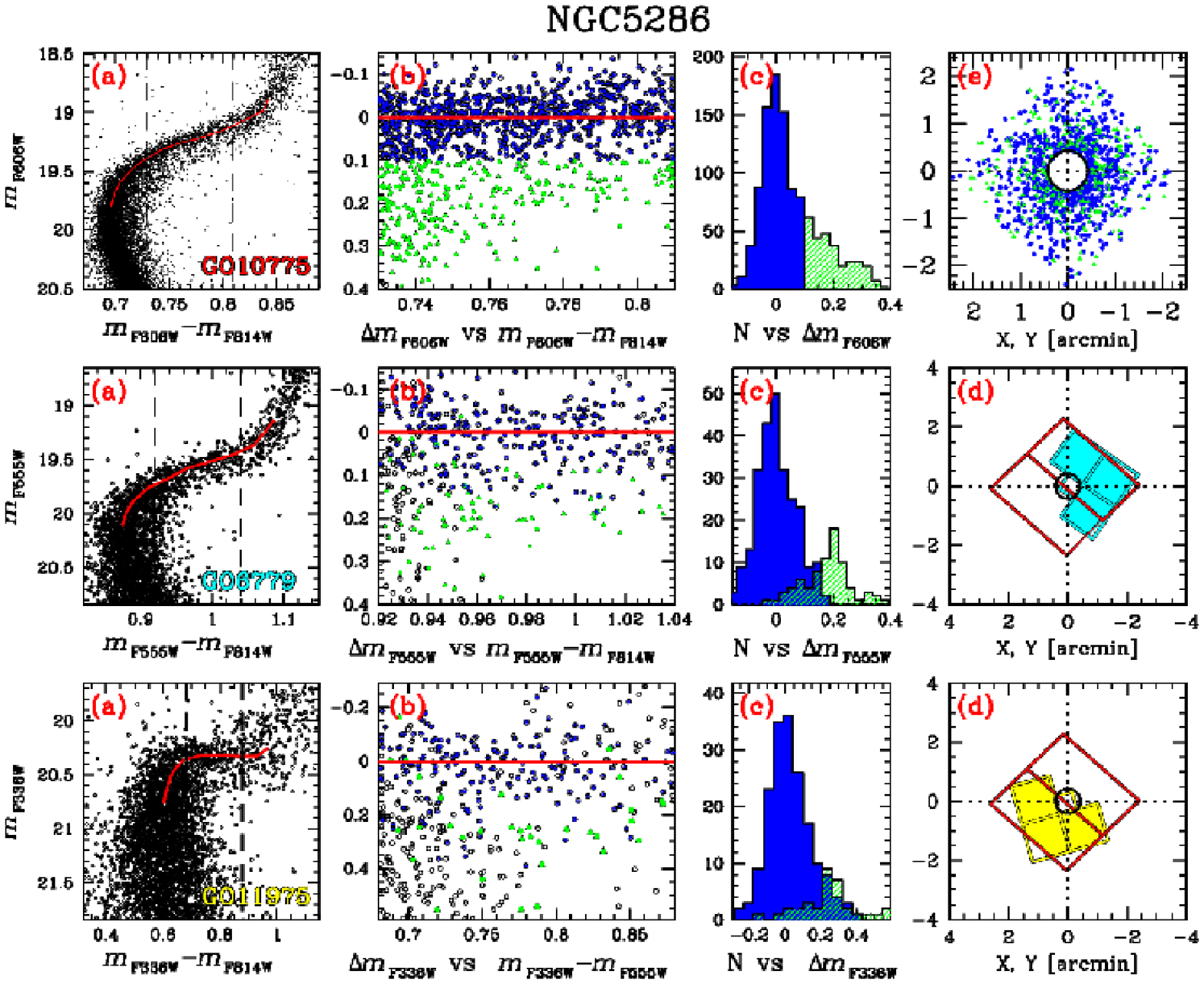}
      \caption{ As in Fig.~\ref{confirmSGB362} for NGC 5286.
	  }
         \label{confirmSGB5286}
   \end{figure}

\begin{figure}[ht!]
   \centering
  \epsscale{.99} 
   \plotone{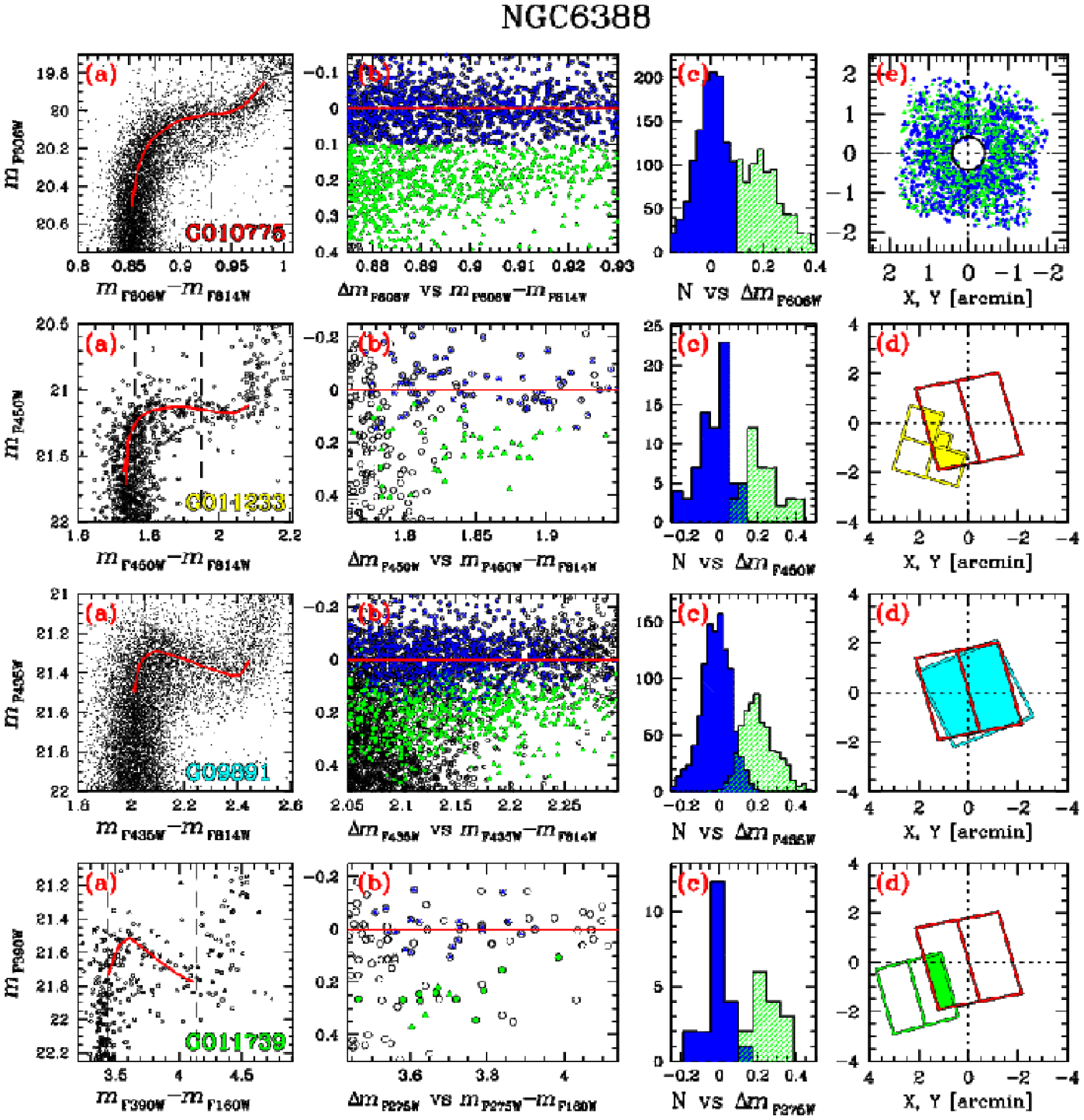}
      \caption{ As in Fig.~\ref{confirmSGB362} for NGC 6388.
	  }
         \label{confirmSGB6388}
   \end{figure}

\begin{figure}[ht!]
   \centering
  \epsscale{.99} 
   \plotone{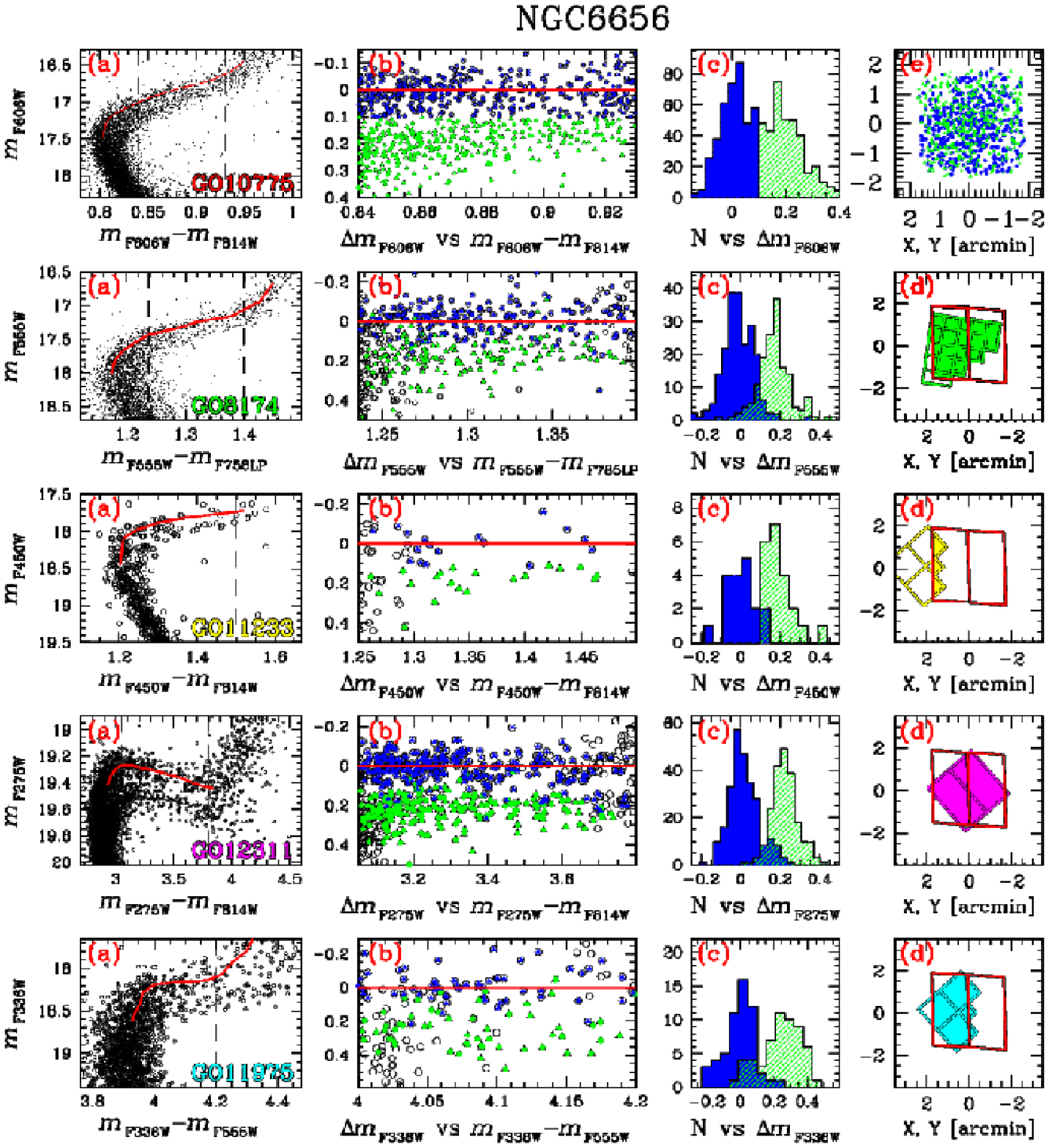}
      \caption{ As in Fig.~\ref{confirmSGB362} for NGC 6656.
	  }
         \label{confirmSGB6656}
   \end{figure}

\begin{figure}[ht!]
   \centering
  \epsscale{.99} 
   \plotone{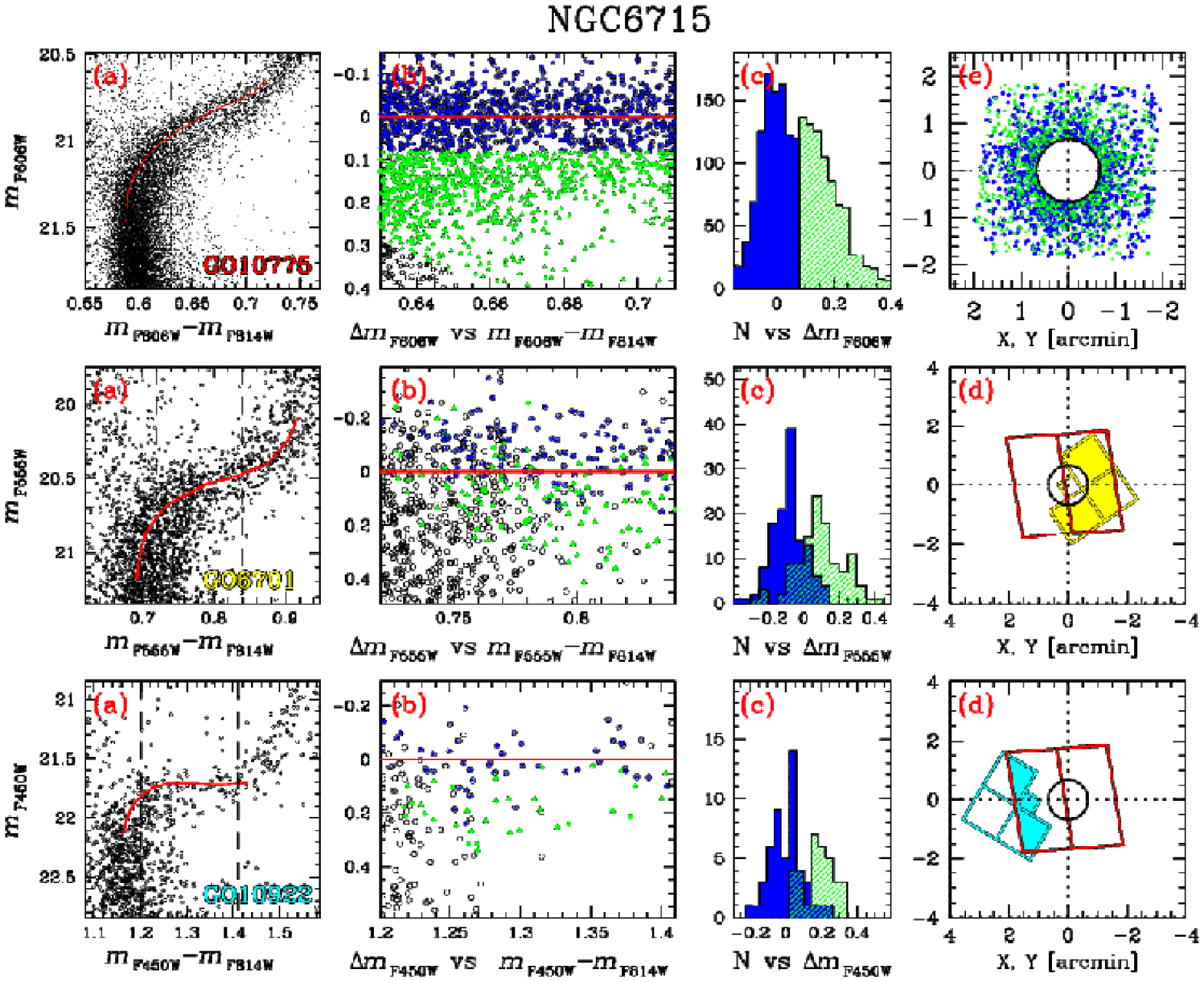}
      \caption{ As in Fig.~\ref{confirmSGB362} for NGC 6715.
	  }
         \label{confirmSGB6715}
   \end{figure}

\begin{figure}[ht!]
   \centering
  \epsscale{.99}
   \plotone{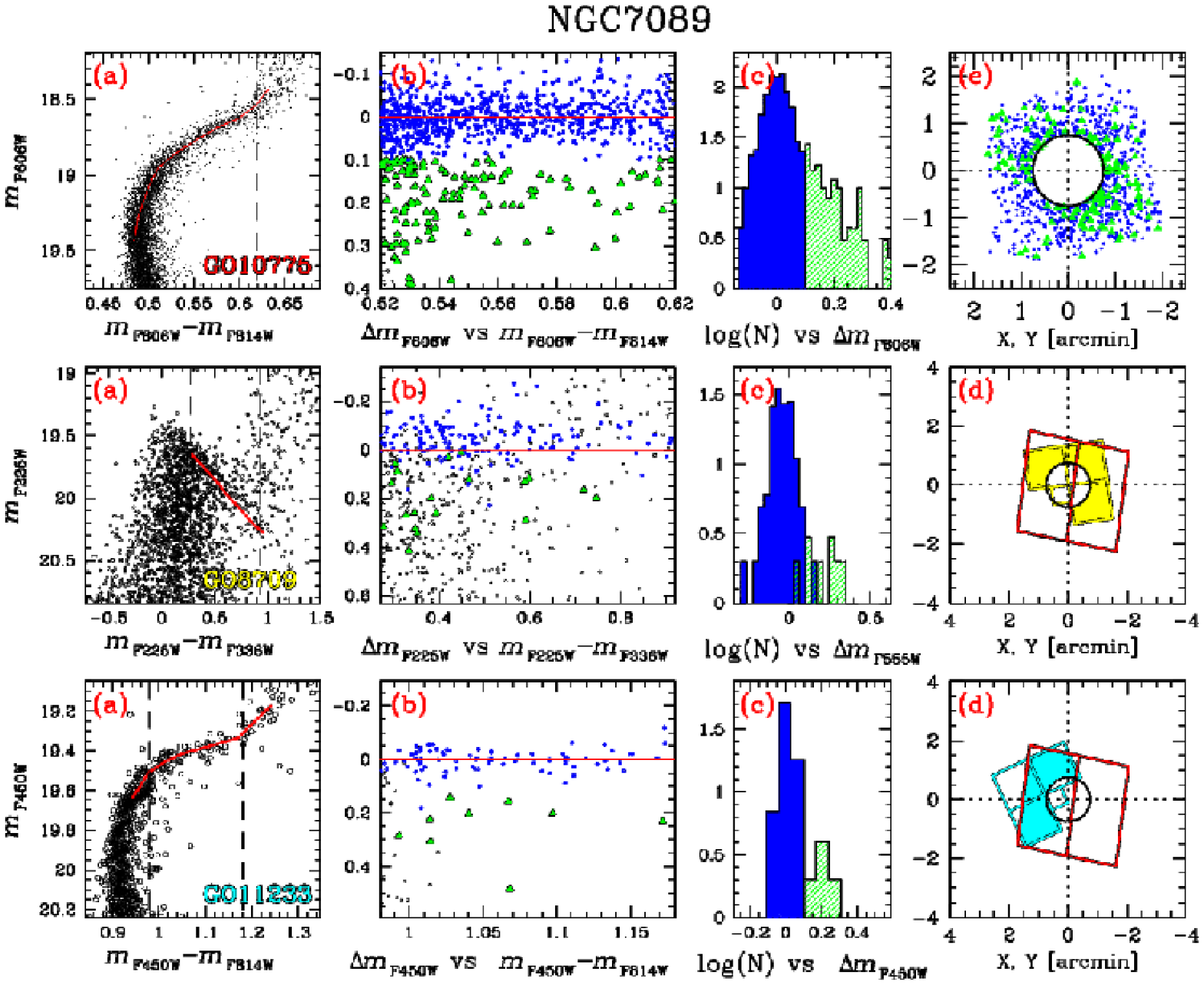}
      \caption{ As in Fig.~\ref{confirmSGB362} for NGC 7089.
	  }
         \label{confirmSGB7089}
   \end{figure}

In this section we make use of a larger set of \textit{HST} photometry
to further confirm and study the SGB spread identified in 
Fig.~\ref{FIGSGB}.  Figures~\ref{confirmSGB362} through \ref{confirmSGB7089} 
present the multi-filter analysis for 
6 out of the eight clusters considered in this work; we describe the 
procedure in detail here for NGC 362.  Panels a of Fig.~\ref{confirmSGB362} 
show the three CMDs of NGC 362 that can be constructed from the three 
independent data sets we have available.  The red line overplotted on 
each CMD is a fiducial line drawn by hand.  We have also calculated the
magnitude difference ($\Delta_{\it m}$) between each star and the 
fiducial line and plotted it (in panels  b) as a function of color,
for stars in the color interval delimited by the dashed lines of 
panels a.

In the top panel b, we divided the SGB as seen in 
F475W-versus-(F435W$-$F625W) from GO-10005 into two groups
based on the F435W magnitude, and colored the bSGB stars blue
and the fSGB stars green. 
 Note that the choice of the two groups of stars is arbitrary, and it is 
not intended to demonstrate that these two groups correspond to two distinct
SGBs. The intent here is to demonstrate that the SGB spread is real and significant,
in all clusters.
In  all the other panels we used
this color-coding for the same stars.
The fact that the green stars lie systematically below the main SGB in all the plots 
demonstrates that they are indeed intrinsically fainter.

The histograms in panels c show the $\Delta_{\it m}$ distribution in
each different CMD.  The fact that the median $\Delta_{\it m}$ 
of fSGB and bSGB stars differs by 7 $\sigma (\Delta_{\it m})$\footnote{ We indicate with $\sigma (\Delta_{\it m})$ the error 
           associated with $\Delta_{\it m}$ as the ratio between 
           the color dispersion measured for all the stars in a 
           given SGB component and the square-root of their number 
           minus one.}
in all the independent data sets demonstrates  that the observed 
spread is real.  Panels d show the footprints of the various
comparison data sets relative to the cluster center.  Only stars 
in the shaded region are used in the analysis presented.  Stars 
within the black circles (where present, for example in NGC\,5286) have poor photometric 
quality because of crowding, and are excluded from our analysis.  
The spatial distribution of bSGB and fSGB stars is shown in the 
uppermost right panel e.  
%

\section{A multi-wavelength analysis of the SGB split/spread }
Recent studies have demonstrated that the our ability to detect
multiple RGB and MS sequences in globular clusters can be very sensitive to 
the particular filter system used to study them.  For example, the RGBs of both 
NGC 6121 and NGC 6752 are clearly bimodal in the {\it U} vs. {\it  U$-$B} CMD,
yet there is no evidence for any split or intrinsic color spread of the RGB 
when they are observed in the {\it B$-$I} color (Marino et al.\ 2008).  Similar 
multiple RGBs have been also observed in CMDs made with the corresponding 
Str\"{o}mgren filters (e.g.\ Grundahl 1999, Yong et al.\ 2008).  These RGB
splits have been interpreted as being related to how CN, CH, and NH bands
affect some spectral regions, but not others  (Marino et al.\ 2008, 
Yong et al.\ 2008).
Similarly, the color differences between the multiple MSs of $\omega$ Centauri,
47 Tucan\ae, NGC 6397, and NGC 6752 depend strongly on the particular color
baseline used to study them.  These MS splits have been attributed to a bimodal distribution in helium abundance.

In this section we take advantage of the huge photometric dataset available 
for each cluster in order to investigate how the SGB morphology changes from 
filter system to filter system.  Although each of our clusters has been observed 
in a different set of bands, each cluster has enough wavelength coverage to make 
direct comparisons possible.

The approach we use here is similar to our study of the MS in $\omega$ Centauri 
(Bellini et al.\ 2010).  Our aim will be to trace out the fiducial sequences 
for the two SGBs in multiple filter bands and quantify the offset between the 
sequences as a function of color at a set of points across the SGB.  If this 
difference is relatively constant with wavelength, then it is indicative of a 
difference in the stellar structure (i.e.\ , radius) for the stars in the two 
sequences.
On the other hand, if the difference varies with wavelength, 
then there could be atmospheric effects going on as well.  

\begin{figure*}[hp!]
   \centering
   \epsscale{.79}
   \plotone{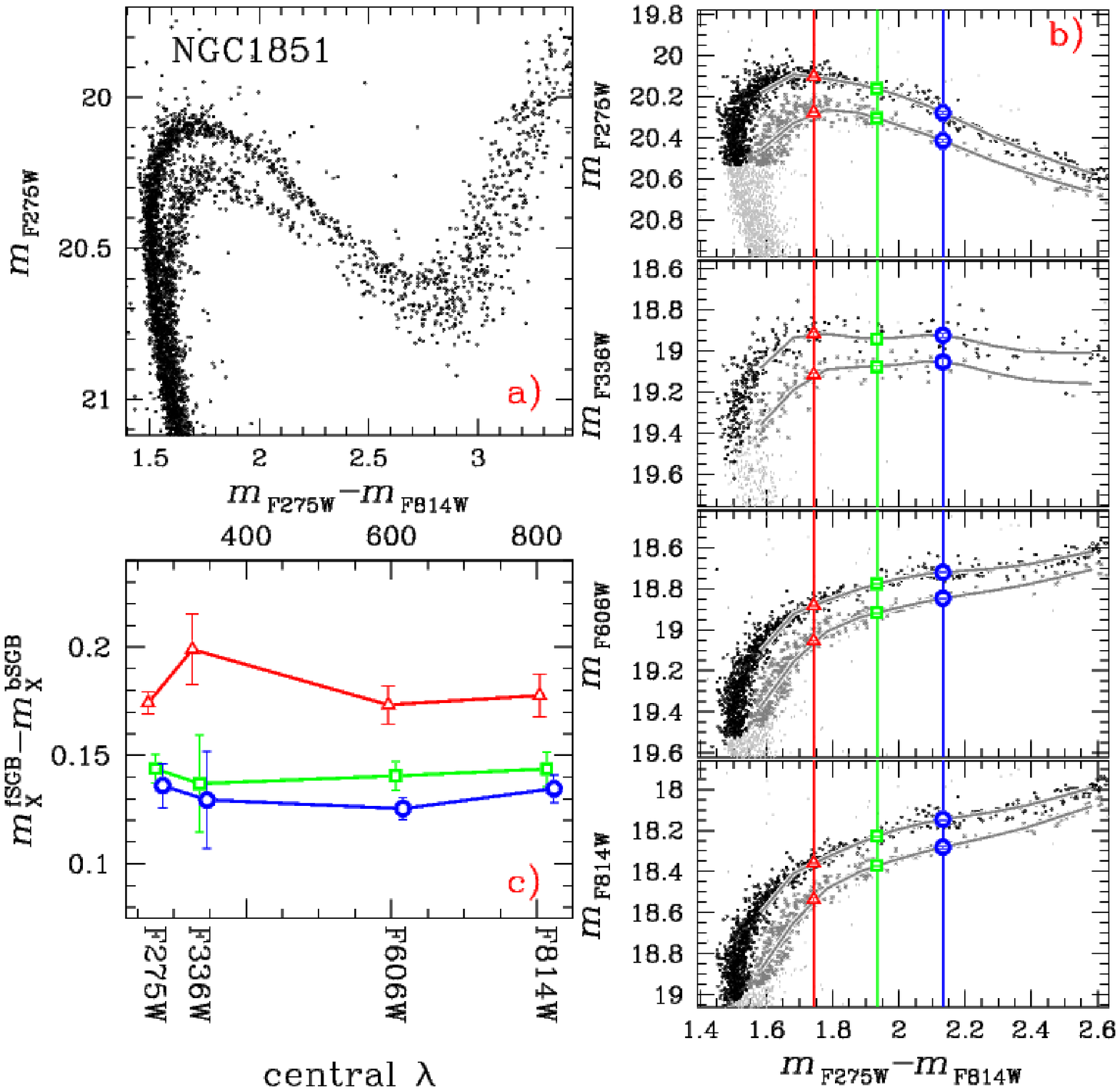}
      \caption{ \textit {Panel a:} ${\it m}_{\rm F275W}$
        versus ${\it m}_{\rm F275W}-{\it m}_{\rm F814W}$ CMD 
        of NGC 1851. \textit{Panel b:} Zoom of the CMDs around the
        SGB, with the sample of fSGB and bSGB (faint and bright SGB)
        colored gray and black respectively. \textit{Panel c:}
        Magnitude difference between the fSGB and the bSGB at 
        ${\it m}_{\rm F275W}-{\it m}_{\rm F814W}=$1.75 (red triangles), 
        1.93 (green squares), and 2.14 (blue triangles). These colors are 
        marked by vertical lines in panel b.
	}
         \label{sgbs1851}
   \end{figure*}

Our approach is illustrated in Fig.~\ref{sgbs1851} for NGC 1851.  For this 
cluster, we have observations for the same stars through the F275W, F336W, 
F606W, and F814W filters.  In the upper left panel (a), we show the stars in 
the color system that best separates the populations.  For this cluster, that 
is in F275W versus F275W$-$F814W. 

We associate each star along the SGB with either the bright or the faint 
population.  In the panels on the right (b), we color-code the stars by 
population and plot them in the same color system (F275W$-$F814W), but with
a different photometric band along the vertical axis in each panel.  The
upper SGB is shown as black points and the lower SGB as medium gray points.
We then trace out the upper and lower sequences by specifying a set of 
fiducial points along each
of them.  

Finally we measured the magnitude difference between the fiducial line of the
faint and the bright sequences corresponding to the three colors indicated by the blue, green, and red line.
We plot the difference between the faint and bright 
sequences  in each photometric band in the lower-left panel (c) for each 
color.
The error bars are indicative of the number and spread of the points that went into the 
value plotted.  It is clear that the separation between the sequences 
is relatively constant with photometric band and it is also relatively 
the same for all three groups.

\begin{figure*}[ht!]
   \centering
  \epsscale{.79}
   \plotone{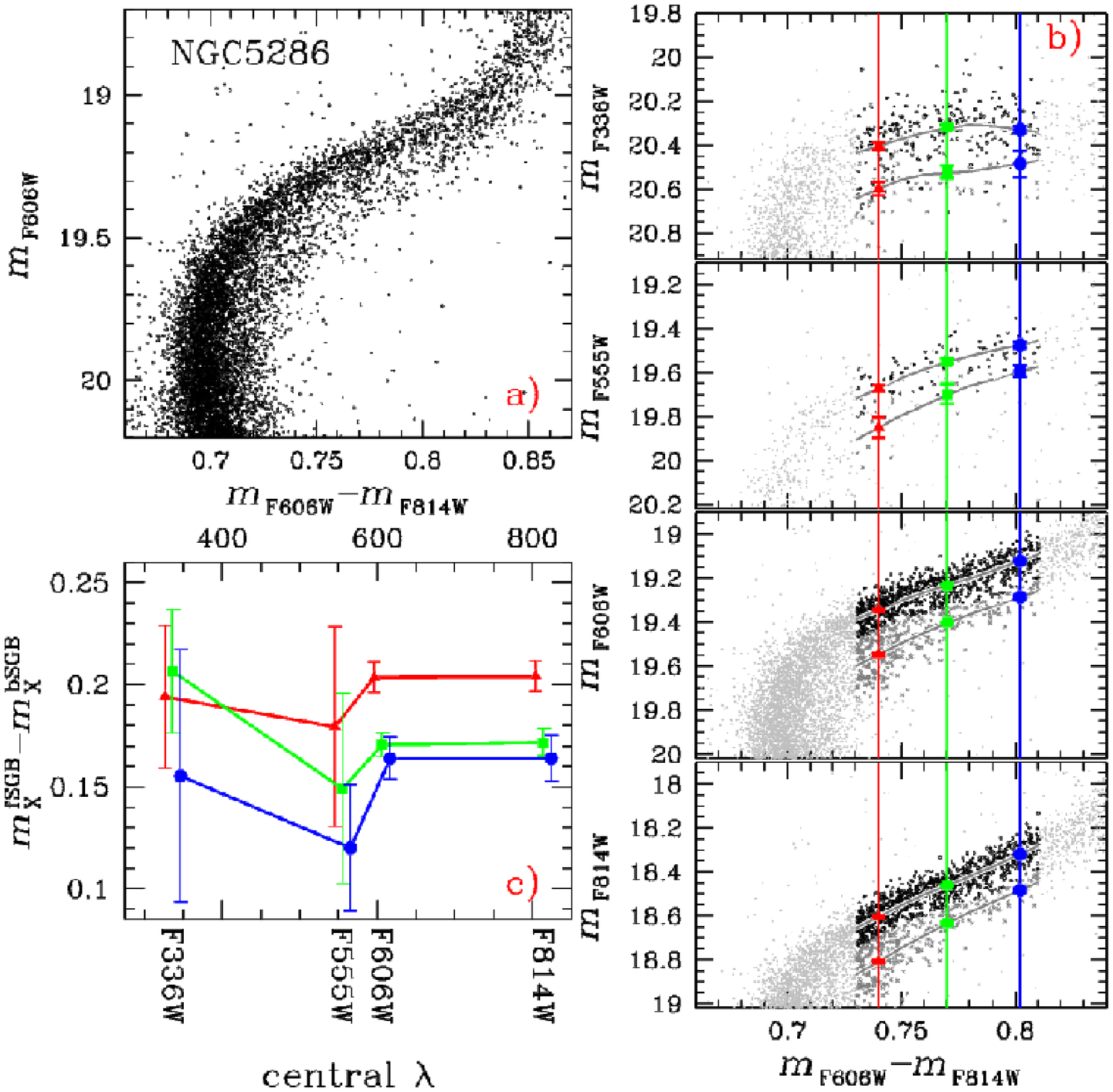}
      \caption{ As in Fig.~\ref{sgbs1851} but for NGC 5286.}
         \label{sgbs5286}
   \end{figure*}

\begin{figure*}[hp!]
   \centering
  \epsscale{.79}
   \plotone{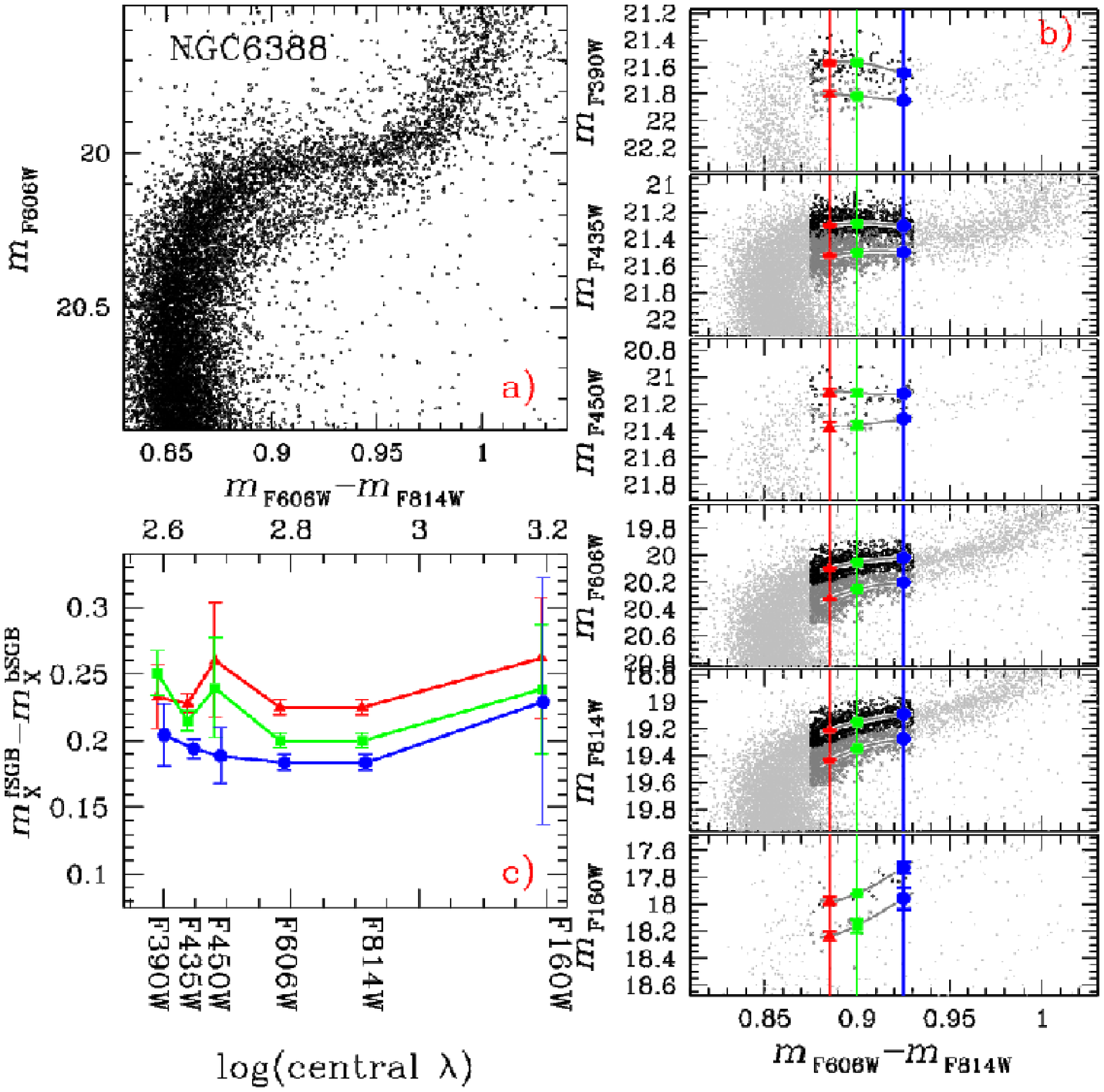}
      \caption{ As in Fig.~\ref{sgbs1851} but for NGC 6388.
	  }
         \label{sgbs6388}
   \end{figure*}

\begin{figure*}[hp!]
   \centering
  \epsscale{.79}
   \plotone{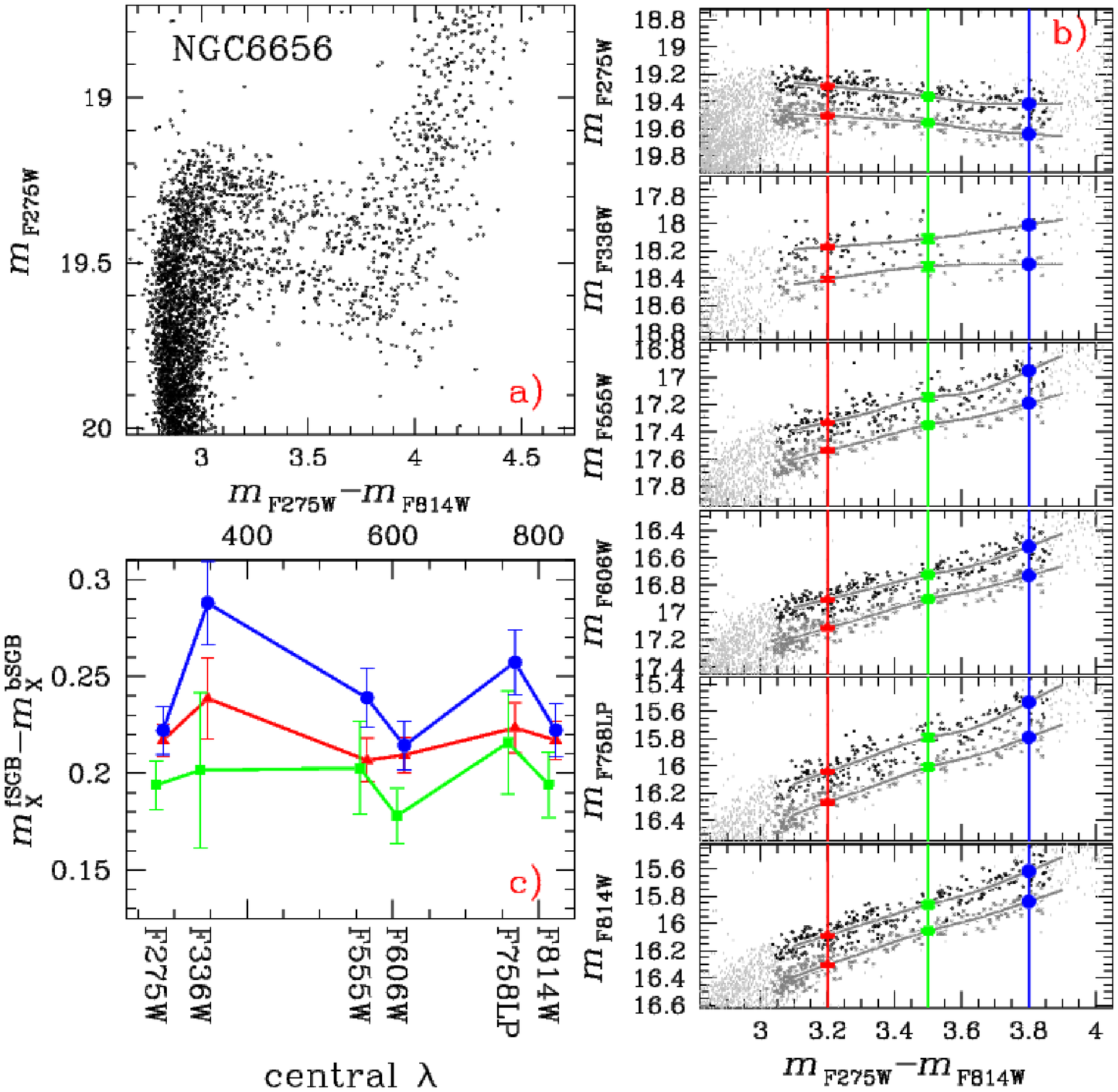}
      \caption{ As in Fig.~\ref{sgbs1851} but for NGC 6656.
	  }
         \label{sgbs6656}
   \end{figure*}

\begin{figure*}[ht!]
   \centering
  \epsscale{.79}
   \plotone{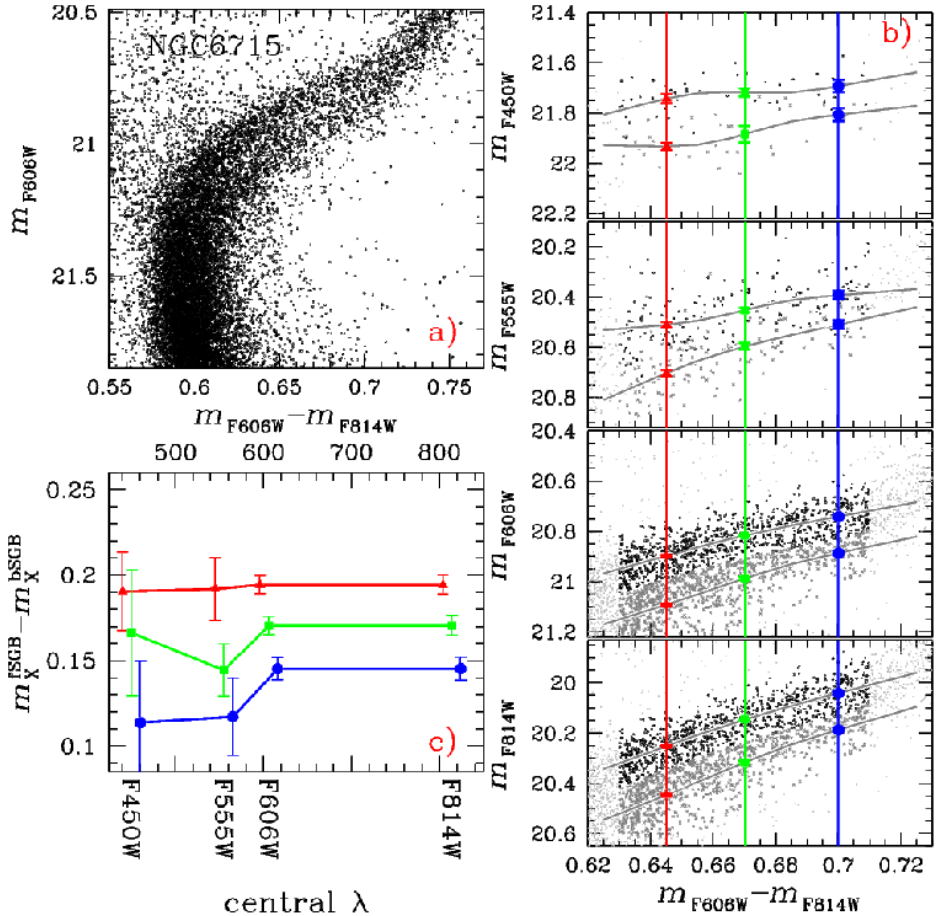}
      \caption{ As in Fig.~\ref{sgbs1851} but for NGC 6715.}
         \label{sgbs6715}
   \end{figure*}

In Figures~\ref{sgbs5286}-\ref{sgbs6715}, we perform the same analysis for   
 NGC 5286, NGC6388, NGC 6656, NGC 6715. We have excluded from this analysis NGC 362 and NGC 7089 because the small number of stars in the fSGB 
does not allow us to obtain a fully realiable fiducial line.
All of these show roughly the same behavior:  the magnitude difference 
between the bSGB and the fSGB has almost no dependence on  wavelength.  This 
behavior rules out the hypothesis that the observed SGB split could be related 
to some effects due to molecular bands on the atmosphere (which would affect only 
specific photometric filters), in contrast to what has been seen on the RGB.
Rather it must be due to some difference in the stellar structure (or mass).  

We will defer a detailed comparison of our CMDs with theoretical isochrones
to a dedicated paper, but it is worth mentioning here that these general 
results are very similar to what has already been found for NGC 1851 by 
Cassisi et al.\ (2008) and Ventura et al.\ (2009). In particular:

\begin{itemize}

\item If we assume that the two stellar populations represented by the 
      two SGBs have the same -- $\alpha$-enhanced -- chemical composition 
      and that the SGB split (or magnitude dispersion)  is due to an age 
      difference alone, stars populating the fSGB should be older by 
      $\sim$1-2 Gyrs.

\item Alternatively, if the bSGB and the fSGB stars have  different chemical 
      compositions, with the fSGB stars being CNO-enriched, the stellar groups 
      corresponding to the two SGBs  could be -- when accounting for the quoted 
      uncertainty on the relative age -- nearly coeval, or the fSGB could be 
      slightly younger.
\end{itemize}
\clearpage
\clearpage
\clearpage
\clearpage

For the clusters analized above
we found that the SGBs can be 
characterized being as bimodal or broad in every filter system.  This 
is in contrast to the well-studied case of 47 Tuc, which has been shown 
to have two close upper SGBs and a more separated lower SGB.  In 47 Tuc, 
we were able to follow the two populations through the entire CMD, 
from the MS to the SGB, RGB, and HB. 

Even though 47 Tuc clearly has a more complicated morphology than
these clusters, since the sequences have been studied both 
photometrically and spectroscopically, it yet may help us understand
the simpler phenomenon at work here.  In 47 Tuc, the two upper SGBs
contain $\sim$92\% of the stars at the center.
The more populous upper sequence (hereafter mSGB)
has been shown to consist of CN-strong/Na-rich/O-poor stars,
while the less populous sequence (uSGB) is made
up of CN-weak/Na-poor/O-rich stars.  
The lowest SGB (named lSGB)
also appears to be CN-strong/Na-rich/O-poor, like the mSGB
(Milone et al.\ 2012a).

\begin{figure*}[hp!]
   \centering
  \epsscale{.79}
   \plotone{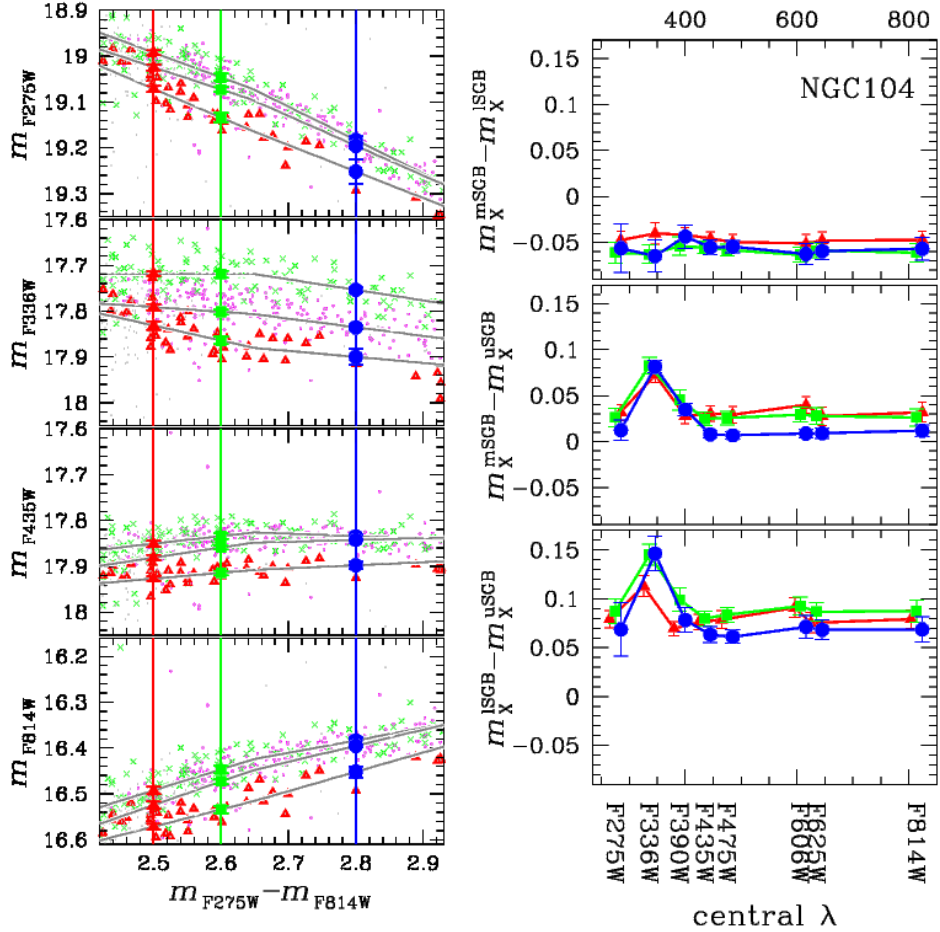}
      \caption{\textit{Left Panel:} Zoom in of the CMDs around the SGB of 47\,Tuc. Green, magenta and red points correspond to uSGB, mSGB, and lSGB stars.  For cleareness we show here only four CMDs.
\textit{Right Panels:} Magnitude difference between the mSGB and the lSGB (upper panel), the mSGB and the uSGB (middle panel), and the lSGB and the uSGB (lower panels). These magnitude differences are calculated at
        ${\it m}_{\rm F275W}-{\it m}_{\rm F814W}=$2.48 (red triangles), 
        2.60 (green squares), and 2.80 (blue triangles). These colors are 
        marked by vertical lines on the left.
        }
         \label{sgbs104}
   \end{figure*}

In order to compare the SGB of 47 Tuc with these clusters, in 
Fig.~\ref{sgbs104} we adapt the procedure introduced in 
Fig.~\ref{sgbs1851} for NGC 1851, to the triple SGB of 47 Tuc.  In the 
CMDs of left panels, the sample of uSGB, mSGB, and lSGB stars 
selected by Milone et al.\ (2012a) are colored green, magenta, and red, 
respectively.  Similar to what was done for NGC 1851, we have calculated 
for each CMD the magnitude difference between the mSGB and the uSGB, 
between the lSGB and the uSGB, and between the lSGB and mSGB 
the  at the color levels indicated by 
the three vertical lines.  In the left panels we again plotted these magnitude 
differences as a function of the central wavelength of the different 
${\it m}_{\rm X}$ bands in our dataset.  We found that the lSGB is 
about 0.05 magnitude fainter than the mSGB, as indicated by the 
filled symbols, and that this magnitude difference is the same, within 
our error bars, at all wavelengths, similar to what we found for the 
other clusters studied in this paper.  The uSGB is typically 
$\sim$0.02 mag brighter than the mSGB
and 0.08 magnitude brighter than the lSGB,
but their magnitude separation 
increases to $\sim$0.08 mag 
and $\sim$0.14 mag in the F336W band (open symbols), respectively. Since 
the F336W filter includes the NH molecular bands at $\lambda$ 
$\sim$3300\AA, we suggest that the latter behavior is due to a 
higher N content (strong NH band absorption) in mSGB stars.

These results support the findings of Di Criscienzo et al.\ (2010) and
Milone et al.\ (2012a). According to these authors,  
the implication is that
the uSGB must correspond to the first stellar population with a chemical composition
similar to the halo stars, while the mSGB should be made by N/Na/He-rich
O/C-poor second-generation stars. Only a small fraction of the second
generation is also characterized by an overall C+N+O increase: these 
stars represent the lSGB.

%
\section{Population ratio}
\label{pratio}
\begin{figure}[hp!]
\centering
\epsscale{.99}
\plotone{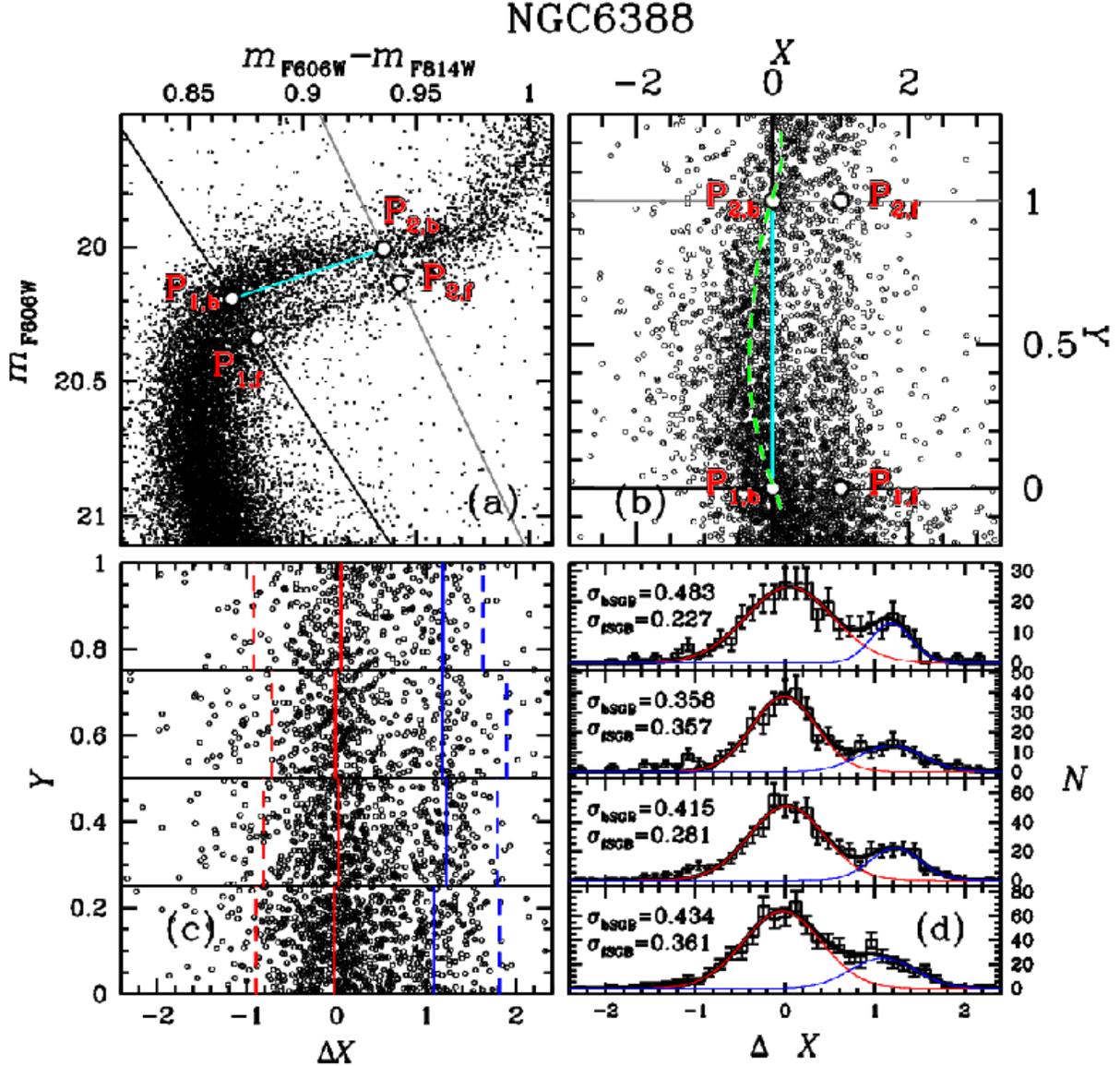}
\caption{Scheme of the procedure used to determine the fraction
         of fSGB and bSGB stars in NGC 6388. Panel a is a zoom of the
         CMD around the SGB; the two lines delimit the SGB portion where
         the split/spread is more evident. In panel b we have transformed the 
         reference frame of panel a and marked with dashed green line the 
         bSGB fiducial drawn by hand.  In panel c we plotted stars between 
         the two continuous lines but after subtracting from the {\it X} 
         value of each star the {\it X} of the fiducial at the corresponding 
         {\it Y}. The four panels d show the $\Delta${\it X} distribution 
         for stars in four $\Delta${\it Y} intervals. The solid lines 
         represent the bi-Gaussian fit. The dispersion of the best-fitting 
         Gaussians are indicated in each box.} \label{Pratio6388}
\end{figure}

\begin{figure}[hp!]
   \centering
  \epsscale{.99}
   \plotone{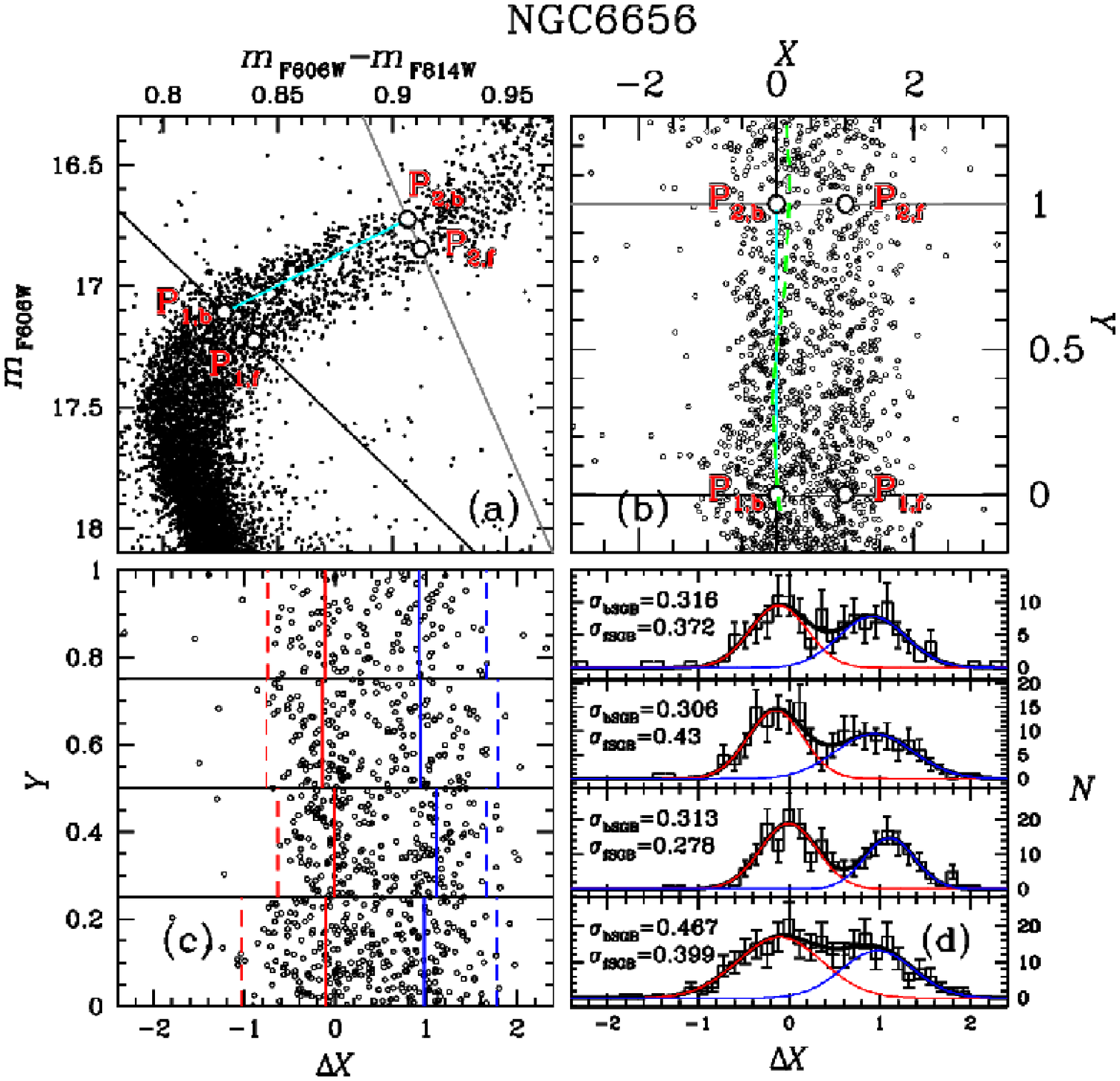}
      \caption{ As in Fig.~\ref{Pratio6388} but for NGC 6656.
	  }
         \label{Pratio6656}
   \end{figure}

\begin{figure}[hp!]
   \centering
  \epsscale{.99}
   \plotone{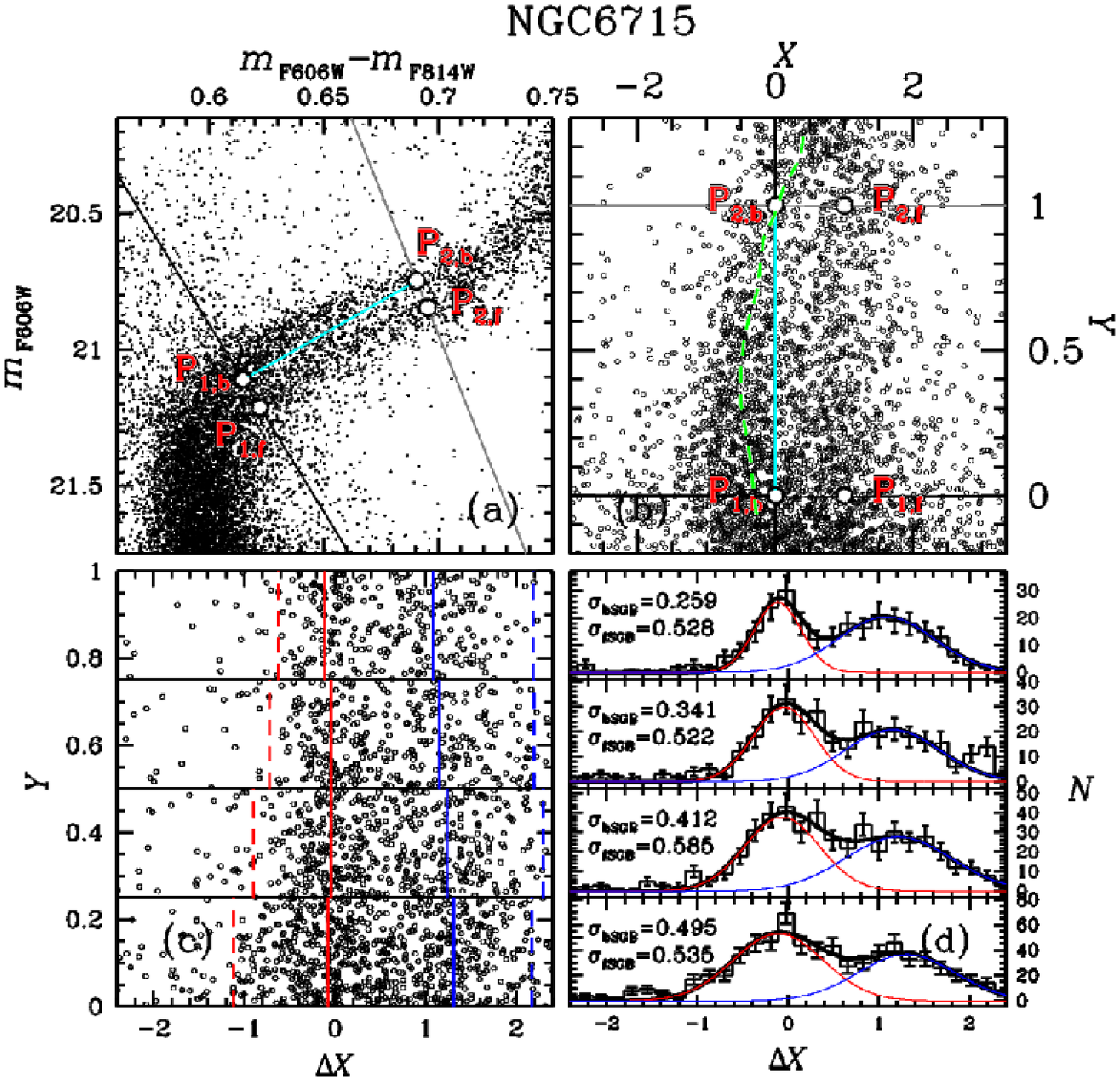}
      \caption{ As in Fig.~\ref{Pratio6388} but for NGC 6715.
	  }
         \label{Pratio6715}
   \end{figure}

\begin{figure}[hp!]
   \centering
  \epsscale{.99}
   \plotone{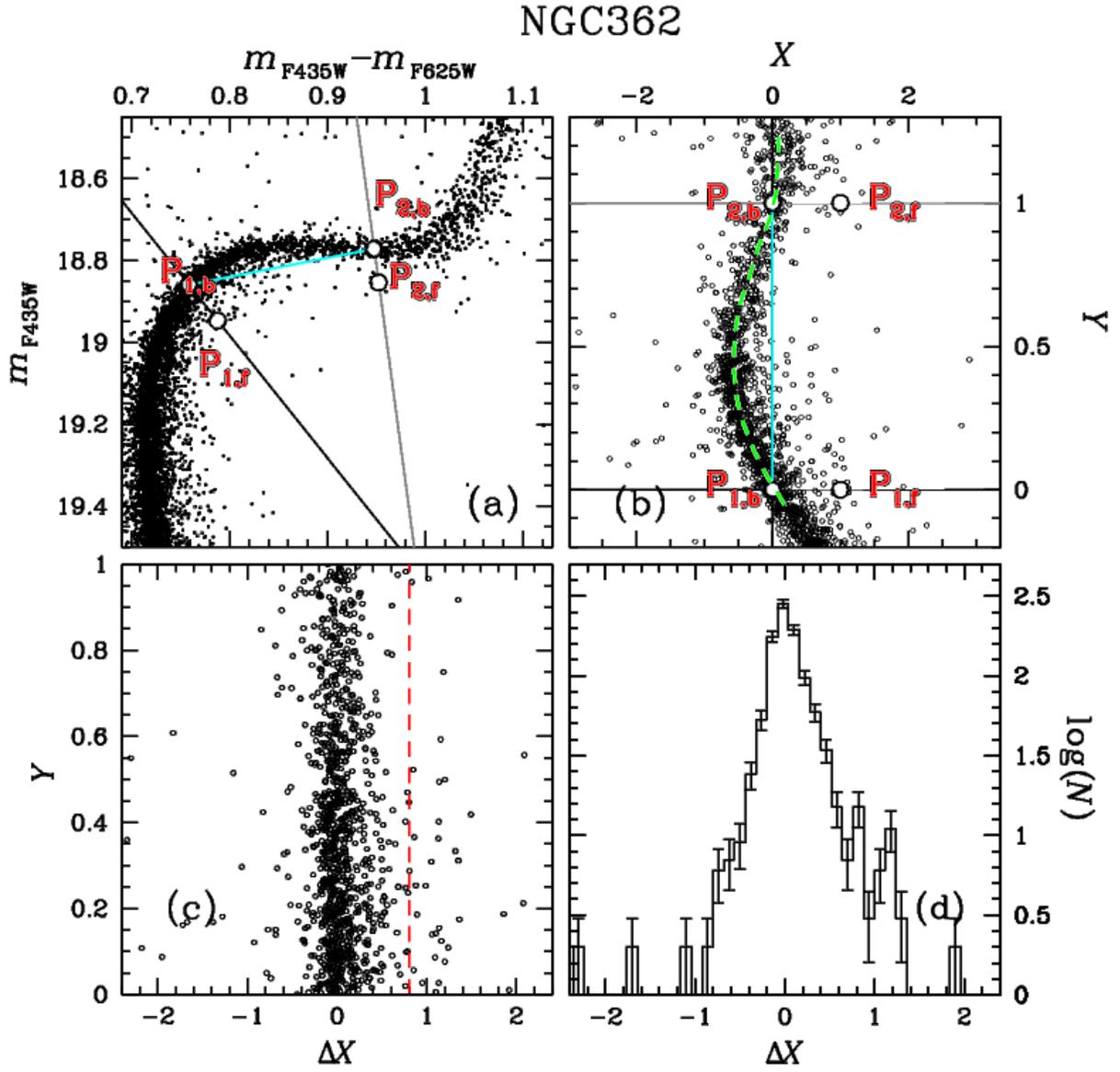}
      \caption{ As in Fig.~\ref{Pratio6388} but for NGC 362.}
         \label{Pratio362}
   \end{figure}

\begin{figure}[hp!]
   \centering
  \epsscale{.99}
   \plotone{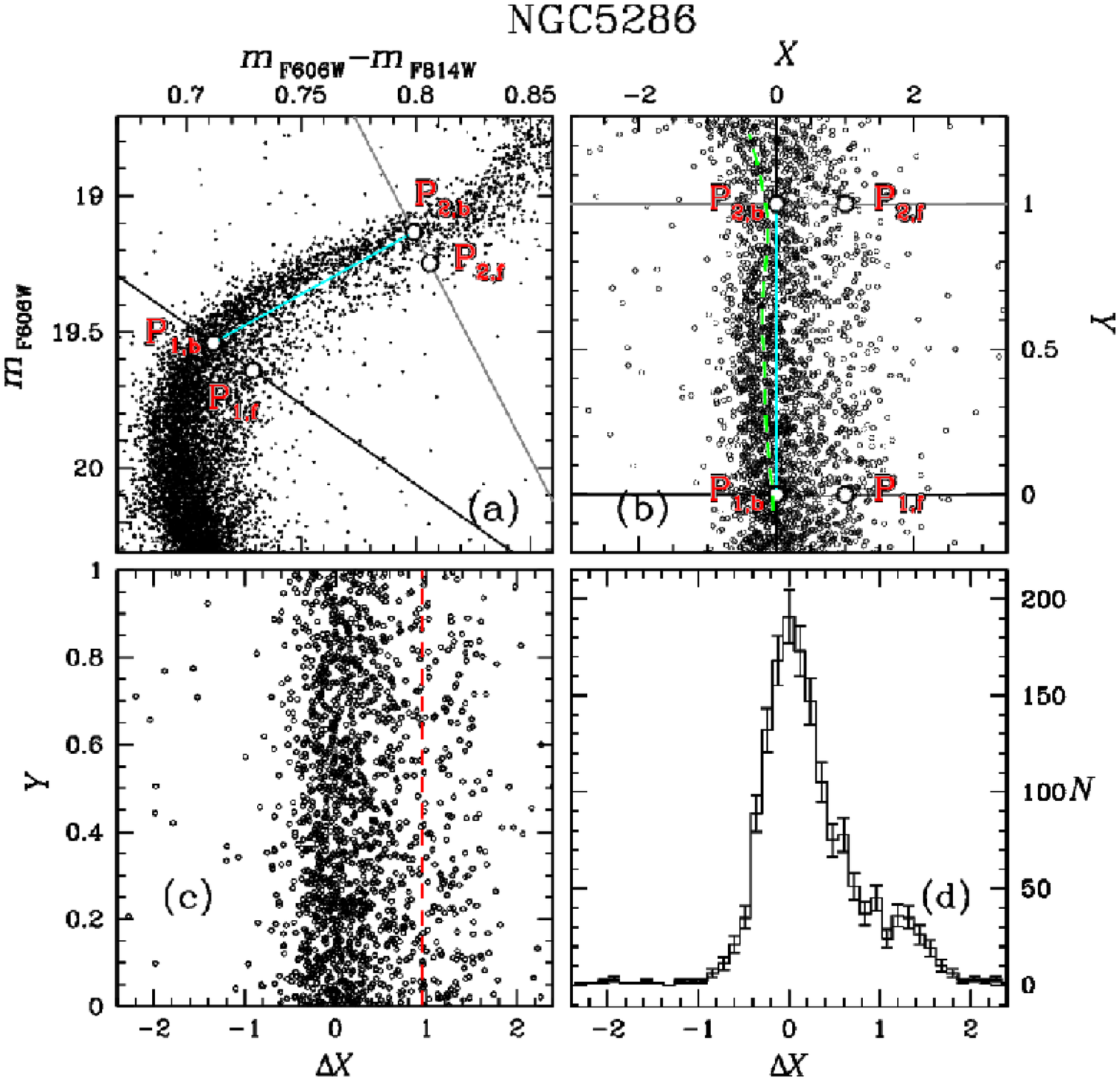}
      \caption{ As in Fig.~\ref{Pratio6388} but for NGC 5286.
	  }
         \label{Pratio5286}
   \end{figure}

\begin{figure}[hp!]
   \centering
  \epsscale{.99}
   \plotone{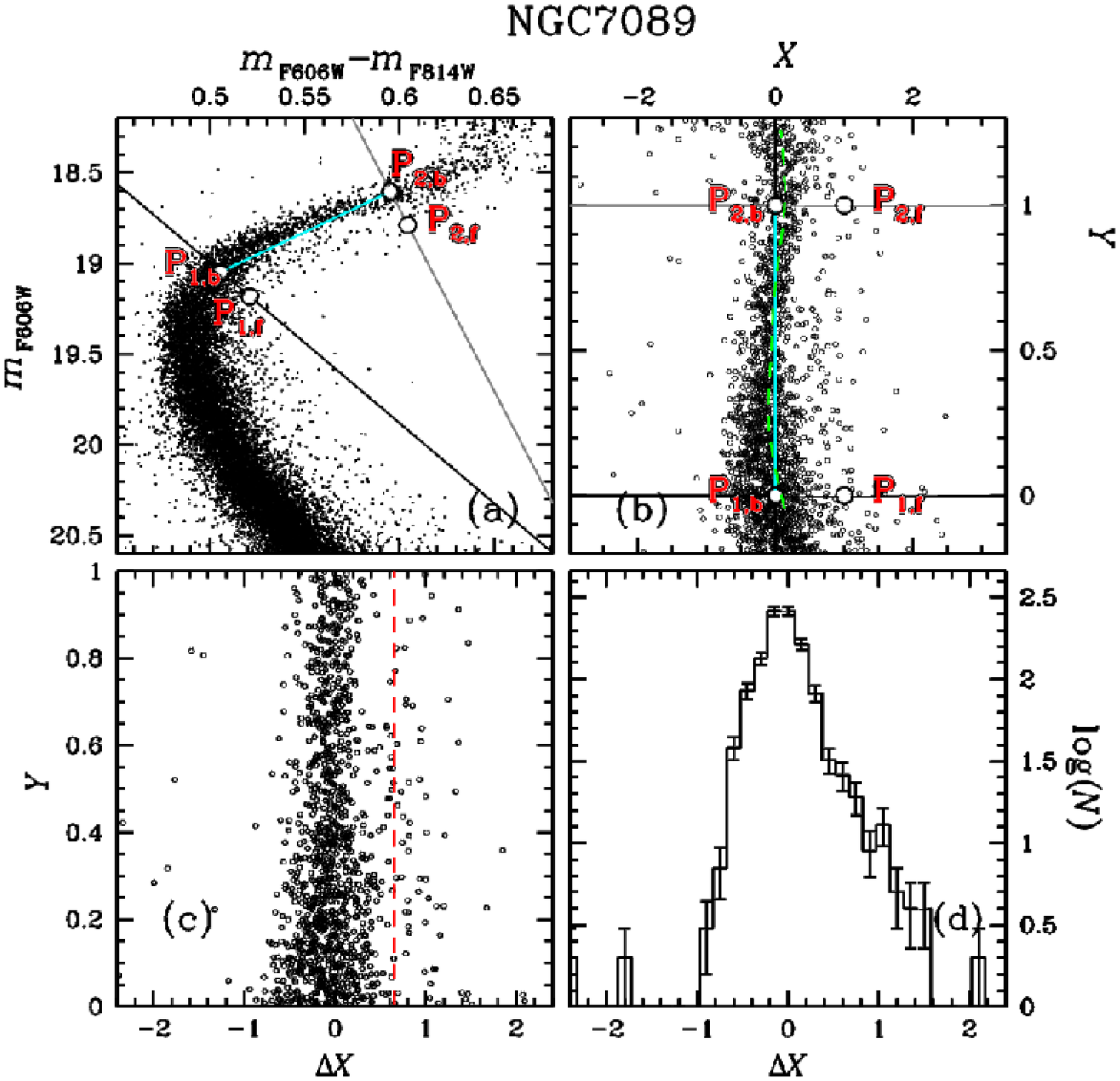}
      \caption{ As in Fig.~\ref{Pratio6388} but for NGC 7089.
	  }
         \label{Pratio7089}
   \end{figure}

The fractions of stars in the different generations can provide fundamental
constraints on the formation and evolution of these stellar systems.
To  estimate the  numbers of fSGB and bSGB stars, we adopted a procedure 
similar to the one used by Milone at al.\ (2009b), and illustrated in 
Figures~\ref{Pratio6388}-~\ref{Pratio7089} for 
six of the 
clusters studied in this paper.

Briefly, we selected by hand two points on the fSGB ($P_{\rm 1,f}$, $P_{\rm 2,f}$) 
and the bSGB ($P_{\rm 1,b}$, $P_{\rm 2,b}$) with the purpose of delimiting 
the SGB region   where the split is most evident, as shown in panel a.  
These 
two points have been connected with a straight line, as shown in panel a.  
Only stars contained within the region between these lines were used in the 
following analysis.  In panel b we have transformed the CMD into a reference 
frame where the origin corresponds to $P_{\rm 1,b}$; $P_{\rm 1,f}$ has 
coordinates (1,0), and the coordinates of $P_{\rm 2,b}$ and $P_{\rm 2,f}$ 
are (0,1) and (1,1), respectively.  
As in Milone et al.\ (2009b), in the 
following, we indicate as `{\it X}' and `{\it Y}' the abscissa and the 
ordinate of this reference frame.  The dashed green line is the fiducial 
line of the bSGB. 

In panel c we have calculated the  difference between the {\it X} value
of each   star  and the  {\it X} of  the    fiducial line ($\Delta$ {\it X}).
In the cases of NGC 6388, NGC 6656, and NGC 6715  the  histograms in panel d 
are the $\Delta${\it X} distribution for  stars  in four  $\Delta${\it Y}  
intervals (see Figs.~\ref{Pratio6388}-\ref{Pratio6715}).  These distributions have been modeled as the sum of two partially 
overlapping  Gaussian functions.  In the cases of NGC 362, NGC 5286, and NGC 7089, 
where the number of fSGB stars is too small not only to fit with an independent Gaussian,  but also to distinguish if the fSGB and the bSGB are two distinct populations or the product of continuous star-formation epishodes.
we have arbitrarily drawn the vertical line in panel c to separate the groups 
of fSGB and bSGB stars (see Figs.~\ref{Pratio362}-\ref{Pratio7089}).  

It is important to note that each of the points
$P_{\rm 1,b}$, $P_{\rm 1,f}$, $P_{\rm 2,b}$, and $P_{\rm 2,f}$---
arbitrarily defined with the sole purpose of isolating a group of stars 
representative of each of the two SGBs---corresponds to a different mass 
($\mathcal{M}_{\rm P1b}$, $\mathcal{M}_{\rm P1f}$,
 $\mathcal{M}_{\rm P2b}$, and $\mathcal{M}_{\rm P2f}$).
To obtain a reliable measure of the fraction  of stars in each of the two 
populations (hereafter: $f_{\rm bSGB}$,  $f_{\rm fSGB}$)  we have to compensate 
for the fact that the two stellar groups that define the two SGBs cover two 
different mass intervals 
     ($\mathcal{M}_{\rm P2f}-\mathcal{M}_{\rm P1f}
              \neq \mathcal{M}_{\rm P2b}-\mathcal{M}_{\rm P1b}$), 
due to the different  evolutionary  lifetimes.   Consequently, the correction 
we have to apply will  be somewhat dependent on the choice of the  mass
function, $\phi({\mathcal M})$.  

To this end, we can calculate the fraction of stars in each branch as:
\begin{center}
$f_{\rm bSGB}  =  \frac { \frac {A_{\rm b}} {N_{\rm b}/N_{\rm f}}}
  {A_{\rm f} +  \frac
    {A_{\rm b}} {N_{\rm b}/N_{\rm f}}} $ \\
$f_{\rm fSGB}  =  \frac {A_{\rm f}} {A_{\rm f} +  \frac {A_{\rm b}}
    {N_{\rm b}/N_{\rm f}}} $
\end{center}
where
$N_{\rm f( b)}=\int_{P_{\rm   1,f(b)}}^{P_{\rm 2,f(b)}} \phi(\mathcal{M})d\mathcal{M}$, 
are the observed numbers of stars, $\phi(\mathcal{M})$ the adopted mass function,  and  $A_{\rm b}$ and $A_{\rm f}$ 
the areas under the Gaussians, for the bSGB and fSGB stars.  
The latter are calculated as the areas of the 
Gaussians that best fit the bSGB and  the fSGB in the cases of NGC 6388, 
NGC 6656, and NGC 6715, while for NGC 362, NGC 5286, and NGC 7089 we assumed 
that $A_{\rm b}$ and $A_{\rm f}$ correspond to the number of stars on the 
blue and the red side of the dashed line of panel c.

As for the  dependence  on  the  adopted mass function,   we ran the following test. We assumed first an heavy-mass-dominated  mass   function ($\alpha=-1.0$) and a then steep ($\alpha=3.0$) mass function. 
Even with  these extreme assumptions, we found  that the mass function effect can change  the relative fSGB/bSGB population ratio by a negligible 3\%. Therefore, for simplicity,  
we adopted a Salpeter (1955) IMF for $\phi(\mathcal{M})$. 

\begin{figure*}[hp!]
   \centering
  \epsscale{.89}
  \plotone{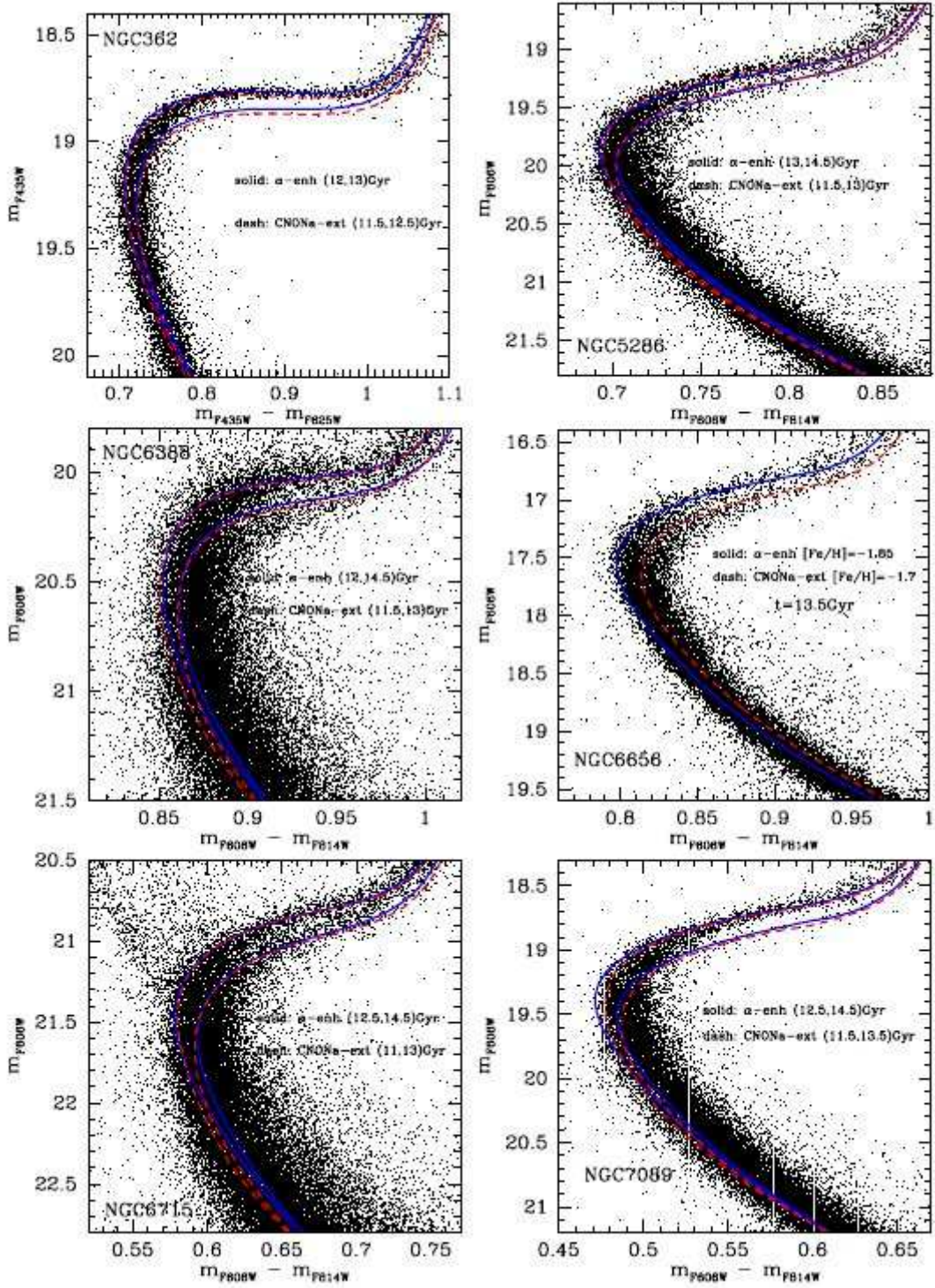}
      \caption{CMD of six clusters with double or spread SGB. The
        dashed and solid lines represent the isochrones for the normal
        population and the extreme CNO-enriched
        ones respectively. Best-fitting ages, distance modulus and
        reddening are quoted in Table~3.}
         \label{iso}
   \end{figure*}

Results are listed in Table~3.
Columns 2 and 3 list the relative number 
of each population obtained by assuming that the two SGBs have the same 
age but  different C+N+O content 
(fSGB stars  with two times more C+N+O than bSGB ones). 
The relative number of bSGB and fSGB 
stars listed in Cols.~4 and 5 are calculated by assuming for the two SGBs the
same overall CNO but different age 
($\Delta$ age $1-2$ Gyr, fSGB older, 
 Fig.~\ref{iso}, see Cassisi et al.\ 2008). 
We found that by 
using the different isochrones the resulting fractions of fSGB and bSGB stars 
changes by less than 0.04.  In most cases these differences are within our 
uncertainties.

We found that in five out of six GCs in Table~3
the bSGB contains 
the majority of the cluster stars, which is also what has been observed in 
NGC 104 (Anderson et al.\ 2009) and NGC 1851 (Milone et al.\ 2009b), bringing 
the statistics to seven out of eight.  The fraction of fSGB and bSGB varies
considerably from cluster to cluster:  it is a few percent in the cases of
NGC 362 and NGC 7089 and $\sim$50\% for the case of NGC 6715.  

\clearpage
\clearpage
\clearpage
\clearpage
\clearpage
\clearpage
%
%
\clearpage
\begin{table}
\centering
\caption{Fraction of fSGB and bSGB stars for the six GCs studied in
         this paper. }
\begin{tabular}{lcccc}
\hline       &  Same age, twice C+N+O &   & Same C+N+O, $\Delta$ age 1-2 Gyr \\ 
\hline  ID   &  $\frac{bSGB}{ALL}$ & $\frac{fSGB}{ALL}$ &  $\frac{bSGB}{ALL}$ & $\frac{fSGB}{ALL}$ \\
\hline
NGC  362  & 0.97$\pm$0.01 & 0.03$\pm$0.01 & 0.99$\pm$0.01 & 0.01$\pm$0.01 \\
NGC 5286  & 0.87$\pm$0.03 & 0.13$\pm$0.03 & 0.85$\pm$0.03 & 0.15$\pm$0.03 \\
NGC 6388  & 0.78$\pm$0.02 & 0.22$\pm$0.02 & 0.90$\pm$0.02 & 0.10$\pm$0.02 \\
NGC 6656  & 0.60$\pm$0.04 & 0.40$\pm$0.04 & 0.56$\pm$0.04 & 0.36$\pm$0.04 \\
NGC 6715  & 0.47$\pm$0.04 & 0.53$\pm$0.04 & 0.49$\pm$0.04 & 0.51$\pm$0.04 \\
NGC 7089  & 0.95$\pm$0.01 & 0.05$\pm$0.01 & 0.97$\pm$0.01 & 0.03$\pm$0.01 \\
\hline
\label{tabellapopratio}
\end{tabular}
\end{table}
%
\section{Summary}
In this paper we have shown that the Galactic GCs NGC 362, NGC 5286, 
NGC 6388, NGC 6656, NGC 6715, and NGC 7089, exhibit double 
 or broadened
SGBs, similar 
to those identified in NGC 1851 (Milone et al.\ 2008) and NGC 104 
(Anderson et al.\ 2009, Milone et al.\ 2012a).  When we compare different CMDs 
of the same cluster made with different magnitudes and color combinations,
we find that the magnitude difference between the bright and the faint SGB 
components remains approximately constant and does not depend on the used filters. 
Therefore, we conclude that 
the split/spread SGB of the GCs studied in this paper can be interpreted in terms of two stellar groups with either a difference in age by $\sim$1-2 Gyr or a large  difference in the total C+N+O abundance, as previously
suggested by Cassisi et al.\ (2008) and Ventura et al.\ (2009) for the case of NGC 1851.  In the cases of NGC 362, NGC 7089, and NGC 5286, the small number of fSGB stars do not allow us to firmly estabilish if the two groups of fSGB and bSGB stars correspond to two distinct stellar populations or if the poorly populated fSGBs are just the tail of an exthended star-formation history.

The fractions of faint 
and bright SGB stars with respect to the total number of SGB stars changes from 
cluster to cluster, and ranges from (0.03:0.97) in the case of NGC 362 to
(0.51:0.49) for the 
Sagittarius dwarf galaxy's GC NGC 6715. 

\begin{acknowledgements}
AB, SC, and GP  acknowledge partial support by the ASI-INAF I/009/10/0 grant, and PRIN-INAF 2010.
GP acknowledge partial support by the Universita' di Padova CPDA101477 grant.
JA acknowledges support from HST grant GO-11233.

\end{acknowledgements}

\bibliographystyle{aa}

\end{document}